\documentclass[debug]{rmaa}

\usepackage{mathtools}
\usepackage{paralist}

\usepackage{psfrag,color}
\usepackage{lineno,hyperref}
\usepackage{comment}
\usepackage{euscript}
\usepackage{array}
\usepackage{paralist}
\usepackage{subfigure}
\usepackage{amsmath}
\usepackage{color}
\usepackage{amssymb}
\usepackage[latin1]{inputenc}
\usepackage{float}
\usepackage[normalem]{ulem}

\usepackage{psfrag,color}

\title{Numerical investigations of the orbital dynamics around a synchronous binary system of asteroids} 

\author{
 Santos, L. B. T.;\altaffilmark{1} 
  de Almeida Jr, Allan Kardec;\altaffilmark{1,2}
  Sousa-Silva, P. A.;\altaffilmark{3} 
  Terra, M. O.;\altaffilmark{4} 
  Sanchez, D. M.;\altaffilmark{5} 
  Aljbaae S. ;\altaffilmark{1} 
  Prado, A. F. B. A.;\altaffilmark{6,7} 
  and Monteiro, F.\altaffilmark{8}}

\altaffiltext{1}{Division of Space Mechanics and Control. National Institute for Space Research, INPE. S\~ao Jos\'e dos Campos, Brazil.}
\altaffiltext{2}{Instituto de Telecomunica\c{c}\~oes, 3810-193 Aveiro, Portugal (allan.junior@inpe.br).}
\altaffiltext{3}{S\~ao Paulo State University, UNESP. S\~ao Jo\~ao da Boa Vista, Brazil. (priscilla.silva@unesp.br).}
\altaffiltext{4}{Technological Institute of Aeronautics, ITA. S\~ao Paulo dos Campos, Brazil. (maisa@ita.br).}
\altaffiltext{5}{The University of Oklahoma, Norman OK, USA.}
\altaffiltext{6}{Postgraduate Division - National Institute for Space Research (INPE), S\~ao Paulo, Brazil.} 
\altaffiltext{7}{Professor, Academy of Engineering, RUDN University, Miklukho-Maklaya street 6, Moscow, Russia, 117198. (antonio.prado@inpe.br).}
\altaffiltext{8}{National Observatory, ON. Rio de Janeiro, Brazil. (filipeastro@on.br).}

\shortauthor{Santos, L. B. T., et al}
\shorttitle{ORBITAL DYNAMICS AROUND A BINARY ASTEROID SYSTEM.}

\listofauthors{L. B. T. dos Santos, de Almeida Jr, Allan Kardec, P. A. Sousa-Silva, Terra, M. O., D. M. Sanchez, Safwan, A.  \& A. F. B. A. Prado}
\indexauthor{Santos, L. B. T.}
\indexauthor{de Almeida Jr, Allan Kardec}
\indexauthor{Sousa-Silva, P. A.}
\indexauthor{Terra, M. O.}
\indexauthor{Sanchez, D. M.}
\indexauthor{Aljbaae, S.}
\indexauthor{Prado, A. F. B. A.}

\abstract{
In this article, equilibrium points and families of periodic orbits in the vicinity of the collinear equilibrium points of a binary asteroid system are investigated with respect to the angular velocity of the secondary body, the mass ratio of the system and the size of the secondary. 
We assume that the gravitational fields of the bodies are modeled  assuming the primary as a mass point and the secondary as a rotating mass dipole. This model allows to compute families of planar and halo periodic orbits that emanate from the equilibrium points $ L_1 $ and $L_2$. The stability and bifurcations of these families are analyzed and the results are compared with the results obtained with the Restricted Three-Body Problem (RTBP). The results provide an overview of the dynamical behavior in the vicinity of a binary asteroid system.}

\resumen{

En este art\'iculo, se investigan los puntos de equilibrio y las familias de \'orbitas peri\'odicas en la vecindad de los puntos de equilibrio colineal de un sistema binario de asteroides con respecto a la velocidad angular del cuerpo secundario, la relaci\'on de masa del sistema y el tama\~no del secundario.

Suponemos que los campos gravitatorios de los cuerpos se modelan asumiendo el primario como punto de masa y el secundario como dipolo de masa giratorio. Este modelo permite calcular familias de \'orbitas planas y halo peri\'odicas que emanan de los puntos de equilibrio $ L_1 $ y $L_2$. Se analiza la estabilidad y bifurcaciones de estas familias y se comparan los resultados con los obtenidos con el Problema de los Tres Cuerpos Restringido (RTBP). Los resultados brindan una descripci\'on general del comportamiento din\'amico en la vecindad de un sistema binario de asteroides.}

\addkeyword{Stability}
\addkeyword{Celestial mechanics}
\addkeyword{Minor planets, asteroids: general}

\begin{document}
\maketitle

\section{Introduction}\label{sec:intro}

In recent years, the investigation and analysis of small celestial bodies have become fundamental to deep space exploration. Thus, understanding the dynamical behavior in the vicinity of small bodies is of great interest for the design of exploration missions and also for planetary science.

However, describing how a particle behaves around these objects is a challenging subject in astrodynamics, mainly due to the combination of the rapid rotation of the asteroids around their axis together with the non-spherical shapes.

In particular, an increasing number of binary asteroid systems has been observed throughout the Solar System and, in particular, among the near-Earth asteroids (NEAs). It is estimated that about 15\% of NEAs larger than 0.3 km are binary systems \citep{2006Icar..181...63P, 2015aste.book..355M}. Most of these binaries are formed by a more massive primary component, usually with nearly spherical shapes, and a small secondary component, generally referred to as satellite \citep{2006Icar..181...63P, 2007Icar..190..250P, 2008Natur.454..188W, 2020AcAau.177...15Z}.

There are several types of binary asteroid systems, which have been grouped according to their physical properties (e.g. size, rotation, mass ratio, diameter) \citep{2007Icar..190..250P}. The characteristics of these groups also suggest different formation mechanisms. As shown by \citet{2007Icar..190..250P}, the Type A binary asteroids are composed of small NEAs, Mars crosses (MC), and Main-Belt Asteroids (MBA), with primary components less than 10 km in diameter and with a component size ratio ($D_s$/$D_p$) less than 0.6. The Type B, in turn, consists of small asteroids with nearly equal size components ($D_s$/$D_p$ $>$ 0.7) and with primary diameters smaller than 20 km. The Types L and W are, respectively, composed of large asteroids ($D$ $>$ 20km) with relatively very small component size ratio ($D_s$/$D_p$ $<$ 0.2) and of small asteroids ($D$ $<$ 20 km) with relatively small satellites ($D_s$/$D_p$ $<$ 0.7) in wide mutual orbits.

Most Type A binary asteroids are synchronous systems, that is, the rotation period of the secondary component is equal to the orbital period around the center of mass of the system \citep{2006Icar..181...63P, 2016Icar267PRAVEC}. Numerical simulations revealed that binary systems are likely to undergo a chaotic process of energy dissipation involving tidal forces that allows the system to evolve to a fully synchronous end state \citep{2011IcarJacobson}. According to \citet{2011IcarJacobson}, the higher the mass ratio of the binary system, the faster the synchronization can be achieved. This happens because each member of the system exerts tidal forces with the same proportion over each other. Thus, as most systems have mass ratios less than 0.5, we find in the literature a larger number of systems with only the secondary component coupled with the orbital movement \citep{2016Icar267PRAVEC}.

Performing semi-analytical and/or numerical investigations of the orbits and equilibrium solutions around asteroid systems using simplified models can be useful to provide some preliminary understanding of such systems \citep{1994CeMDA..59..253W, 2011Ap&SS.333..409L}. 
Simplified models can be used to approximate the gravitational field to irregularly shaped bodies, requiring less computational effort and generating considerable results in a short period of time.
Another advantage of using a simplified model is that we can easily investigate the effects of a given parameter on the dynamics of a spacecraft around asteroids, such as, the distribution of stable periodic orbits \citep{2017Ap&SS.362..169L}, the stability of the equilibrium points \citep{2015Ap&SS.356...29Z, 2017ApSS.362...61B}, as well as the permissible parking regions \citep{2015RAA....15.1571Y, 2016JGCD...39.1223Z}. In addition, simplified models can be used to support the orbit design \citep {2017Ap&SS.362..229W} and feedback control \citep{2017Ap&SS.362...27Y}. 

Due to their advantage and considerable results, several simplified models have been used proposed to study the orbital dynamics of a particle in the vicinity of irregular bodies. For example, \citet{1999imda.coll..169R, 2001CeMDA..81..235R} analyzed the dynamics of a particle under the gravitational force of an asteroid modeled as a straight segment. \citet{2016Ap&SS.361...14Z} analyzed the influence of the parameters $k$ (angular velocity) and $\mu$ (mass ratio) in the equilibrium solutions using the rotating mass dipole model and observed that there are always 5 equilibrium points when considering the primary bodies as points of mass. Other works have investigated the dynamics around small irregular bodies using a simplified model given by an homogeneous cube \citep{2011Ap&SS.333..409L}, a simple flat plate \citep{Blesa}, a rotating mass dipole \citep{2015Ap&SS.356...29Z, 2017ApSS.362...61B, 2017Ap&SS.362..202D}, the dipole segment model \citep{2018AJ....155...85Z}, a rotating mass tripole \citep{2017Ap&SS.362..169L, leotripole, leotripole3d}, and many others. 

In particular, aiming to understand the dynamical environment in the vicinity of irregular bodies, \citet{2020MNRAS.496.1645A} investigated the dynamics of a spacecraft around the asynchronous equal-mass binary asteroid (90) Antiope, the authors applied the Mascon gravity framework using the shaped polyhedral source \citep{chanut_2015a, aljbaae_2017} to consider the perturbation due to the polyhedral shape of the components. The perturbations of the solar radiation pressure at the perihelion and aphelion distances of the asteroid from the Sun is also considered in that study. In order to investigate the stability of periodic orbits, \citep{Chappaz} considered the asynchronous binary asteroid system using the triaxial ellipsoid model and observed that the non-spherical shape of the secondary body significantly influences the behavior of the halo orbit around $ L_1 $ and $ L_2 $ .

As said before, simplified models are useful to provide some preliminary understanding of the motion around binary systems, and the circular restricted three-body problem is suitable and often used to investigate the dynamics around small bodies \citep{dealmeidajrbert22}. Furthermore, even landing trajectories has been evaluated using a spherical shape for the gravitational field of the primaries in the circular restricted three-body problem \citep{TardivelScheeres2013,Celik2017,ferrarilav2016}. Although the orbit-attitude coupled equations of motion for a bynary asteroid can be obtained using a more sophisticated model, which takes into consideration a potential for a non-spherical distribution of mass \citep{ScheeresBAresi2019,WenZeng}, they are only essentials for very close encounters, such as for landing approaches. In this study, the dynamics is investigated for orbits around the binary system of asteroids.
Thus, in this contribution, a more simplified model is used, whose results capture the essentials parts of the physics of the problem, although its accuracy depends on the parameters of the specific mission.
Therefore, we carry out a numerical investigation using the simplified model called a Restricted Synchronous Three-Body Problem, as introduced by \citep{2017ApSS.362...61B}. The practical advantage of using this model is that we can, in a relatively simple way, analyze the influence of the dimension of the secondary body on the dynamics of a spacecraft in the neighborhood of $M_2$.

We focus on the behavior of a particle of negligible mass in the vicinity of a binary system of type A small bodies (NEAs and MBAs). The reason for choosing this class of asteroids is that, the NEAs, in particular, are asteroids that pass near the Earth and most of the systems that are part of this class are synchronous systems. Our aim is to understand how the parameters of the dipole, dimension ($ d $) and the mass ratio ($ \mu^* $) of the system, influence the stability, period and, bifurcation of the periodic orbits around the equilibrium points. In Section \ref{equation} we provide the equation of motion of the three-body synchronous restricted problem. In Section \ref{CEP}, we investigate the influence of the force ratio ($k$) on the appearance of the equilibrium points, keeping the values of $\mu^*$ and $d$ fixed. Then, in Section \ref{POAFSC}, we investigate the influence of $\mu^*$ and $d$ on periodic orbits (planar and halo) around the equilibrium points $L_1$ and $L_2$, considering $k$ fixed ($k$ = 1). Finally, in Section \ref{conclusion}, we provide the final considerations that were obtained in this article.

\section{Equations of motion}
\label{equation}

\begin{figure}[!htbp]
\centering\includegraphics[scale=0.9]{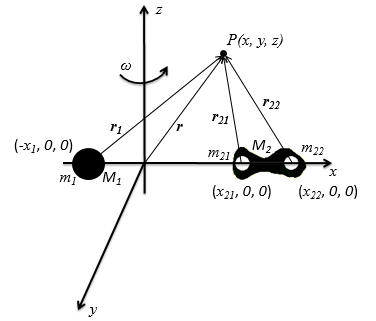}
 \caption{Representative image of the geometric shape of the system considered (out of scale).}
    \label{geometric shape}
\end{figure}

Consider that the motion of a particle with negligible mass, $ P (x,~ y, ~z) $, is dominated by the gravitational forces of the primary bodies $ M_{1} $ and $ M_{2} $. As already mentioned, the distance between $ M_{1} $ and $M_2$ is assumed to be $D = 12$ km, which will be the normalization factor in the rest of this work. The larger primary is considered to be a point mass with mass $ m_1 $ and the secondary is modeled as a rotating mass dipole formed by $ m_{21} $ and $ m_{22} $, as shown in Figure \ref{geometric shape}.

In canonical units, the sum of the masses of the bodies $ M_1 $ and $ M_2 $ is unitary. In this work, for all numerical simulations, we assume that $ m_{1} > m_{21} = m_{22} $ and that the mass ratio is defined by $ \mu^{*} = m_{21}/(m_{1} + m_{21} + m_{22}) $. By analogy, $ \mu^{*} = \mu/2 $, with $ \mu $ being the usual mass ratio used in the classical restricted three-body problem.

The angular velocity, given by  $\boldsymbol\omega$ = $\omega\textbf{z}$, is aligned with the $z$-axis of the system.
Here, the unit of time is defined such that the orbital period of the primary bodies around the center of mass of the system is equal to $ \omega^{-1} $. Because the system is synchronous, the orbital period of $M_2$ around the center of mass is the same as its orbital period around the axis of the dipole. 

With respect to the barycentric rotating frame, the masses $ m_{1} $, $ m_{21} $ and $ m_{22}$ are fixed along the $x$-axis with coordinates $ x_{1} = -2\mu^* $, $ x_{21} = -2\mu^* - \frac{d}{2} + 1 $ and $ x_{22} = -2\mu^* + \frac{d}{2} + 1 $, respectively, where $d$, given in canonical units, is the distance between $m_{21} $ and $ m_{22} $. 

Using the generalized potential 
\begin{equation}
\Omega = \frac{x^2+y^2}{2} + k\left (\frac{1-2\mu^*}{r_1}+\frac{\mu^*}{r_{21}}+\frac{\mu^*}{r_{22}} \right ),
\label{Omega}
\end{equation}
we can write the equations of motion of $P$ in a rotating frame centered on the barycenter of the system ($M_1$-$M_2$) as follows:
\begin{equation}
\begin{bmatrix}
\dot{x}\\
\dot{y}\\
\dot{z}\\
\ddot{x}\\
\ddot{y}\\
\ddot{z}
\end{bmatrix}
=
\begin{bmatrix}
\dot{x}\\
\dot{y}\\
\dot{z}\\  
2\dot{y} + D_x\Omega\\
-2\dot{x} + D_y\Omega\\
D_z\Omega
\end{bmatrix}, 
\label{seila2}
\end{equation}

\noindent
with
\begin{equation*}
\begin{split}
r_1 = \sqrt{(x-x_1)^2 + y^2+z^2},\\
r_{21} = \sqrt{(x-x_{21})^2 + y^2+z^2},\\
r_{22} = \sqrt{(x-x_{22})^2 + y^2+z^2},
\label{distancias}
\end{split}
\end{equation*}
where $D_x\Omega$ denotes the partial derivative of $ \Omega $ with respect to $ x $ and the same notation is used for $ y $ and $ z $. 
The dimensionless parameter $k$ represents the ratio between gravitational and centrifugal accelerations, $ k = G(M)/(\omega^{*2}D^3) $, where $ M $ is the total mass of the system in kg, $ \omega^{*} $ is the angular velocity of the $M_2$ in rad/s, $ D $ is the distance, in meters, between $ M_1 $ and the center of mass of $ M_2 $ and, finally, $G$ = $6.67408\times 10^{-11} m^3 kg^{-1} s^{-2}$ \citep{2015Ap&SS.356...29Z, 2016AdSpR..58..387F}.

The free parameters of the system are $d$, $\mu^*$ and $k$, which correspond, respectively, to the size of $ M_2 $, the mass ratio and a parameter accounting for the rotation of the asteroid. When $ k $ is equal to 1, the bodies orbit the center of mass of the system without any internal forces in the dipole. On the other hand, when $ k ~ <~ 1 $, the dipole is stretching, while it is compressing when $k ~> ~1$. Therefore, depending on the class of the binary system being analyzed, we need to consider the force ratio value ($k$). 
A particular case occurs when $d$ (distance from the mass dipole) is equal to zero, causing the bodies of mass $m_{21}$ and $m_{22}$ to overlap, becoming a point of mass, with mass ratio 2$\mu^*$.
The classical Restricted Three-Body Problem corresponds to the particular case $d=0$ and $k=1$ \citep{mc, 1967torp.book.....S}. Also, when $d$ $\neq$ 0 and $k$ = 1, we have the Restricted Synchronous Three-Body Problem \citep{2017ApSS.362...61B}. 
\subsection{Equilibrium point and stability analysis}
\label{sec:Analysis stability}

Let $\mathbf{x}$ = $(x, ~y, ~z, ~\dot{x}, ~\dot{y}, ~\dot{z})$ $\in$ $\mathbb{R}^6$ be the state vector of a massless particle and $f: \mathbb{R}^6 ~\rightarrow ~\mathbb{R}^6$ be
\begin{equation}
f(\mathbf{x})
=
\begin{bmatrix}
f_1\\
f_2\\
f_3\\
f_4\\
f_5\\
f_6
\end{bmatrix}
=
\begin{bmatrix}
\dot{x}\\
\dot{y}\\
\dot{z}\\
2\dot{y} + D_x\Omega\\
-2\dot{x} + D_y\Omega\\
D_z\Omega
\end{bmatrix}.
\label{seila3}
\end{equation}

\noindent

The equilibrium points $L_i$, $i=1,2,3,4,5,$ are defined as the zeros of $ f (\mathbf{x}) $. To determine the linear stability of each equilibrium, one needs to translate the origin to the position of this equilibrium point and linearize the equations of motions around this point. Thus, the linearization over any of these equilibrium points is
\begin{equation}
 \dot{\mathbf{x}} = D_{L_i}\mathbf{x}
\label{seila4}
\end{equation}
where $D_{L_i}$ is the derivative of $ f (\mathbf{x}) $ computed at the equilibrium point $L_i$.

To determine the linear stability of the equilibrium points $ (L_i $, $ i = 1,2,3,4 $ and $ 5 $), it is necessary to transfer the origin of the coordinate system to the position of the equilibrium points $ ({\it x_0, y_0, z_0}) $ and then linearize the equations of motion around these points, obtaining the results shown below.
\begin{multline}
\ddot{\xi} - 2\dot{\eta} = \Omega_{xx}(x_0,y_0,z_0)\xi + \Omega_{xy}(x_0,y_0,z_0)\eta +\Omega_{xz}(x_0,y_0,z_0)\zeta 
\label{AEPxx}
\end{multline}

\begin{multline}
\ddot{\eta} + 2\dot{\xi} = \Omega_{yx}(x_0,y_0,z_0)\xi + \Omega_{yy}(x_0,y_0,z_0)\eta +\Omega_{yz}(x_0,y_0,z_0)\zeta
\label{AEPyy}
\end{multline}

\begin{multline}
\ddot{\zeta}  = \Omega_{zx}(x_0,y_0,z_0)\xi + \Omega_{zy}(x_0,y_0,z_0)\eta +\Omega_{zz}(x_0,y_0,z_0)\zeta
\label{AEPzz}
\end{multline}
where the partial derivatives in $ (x_0, y_0, z_0) $ mean that the value is calculated at the equilibrium point being analyzed. Partial derivatives are shown in Equations \ref{XX} - \ref{YZ}.

\begin{multline}
\Omega_{xx}= 1+k\bigg[ \bigg. \frac{3(1-2\mu^*)(x-x_1)^2}{((x-x_1)^2+y^2+z^2)^{5/2}} - \frac{1-2\mu^*}{((x-x_1)^2+y^2+z^2)^{3/2})}+ \\  \frac{3\mu^*(x-x_{21})^2}{((x-x_{21})^2+y^2+z^2)^{5/2}}- \frac{\mu^*}{((x-x_{21})^2+y^2+z^2)^{3/2}} - \\ \frac{3\mu^*(x-x_{22})^2}{((x-x_{22})^2+y^2+z^2)^{5/2}} + 
\frac{\mu^*}{((x-x_{22})^2+y^2+z^2)^{3/2}} \bigg. \bigg],
\label{XX}
\end{multline}

\begin{multline}
\Omega_{yy}= 1+k\bigg[ \bigg.
\frac{3(1-2\mu^*)y^2}{((x-x_1)^2+y^2+z^2)^{5/2}} - \frac{1-2\mu^*}{((x-x_1)^2+y^2+z^2)^{3/2}}  + \\ \frac{3\mu^*y^2}{((x-x_{21})^2+y^2+z^2)^{5/2}}- \frac{\mu^*}{((x-x_{21})^2+y^2+z^2)^{3/2}}  + \\ \frac{3\mu^*y^2}{((x-x_{22})^2+y^2+z^2)^{5/2}} - 
\frac{\mu^*}{((x-x_{22})^2+y^2+z^2)^{3/2}}\bigg. \bigg],
\label{YY}
\end{multline}

\begin{multline}
\Omega_{zz}= k\bigg[ \bigg.\frac{3(1-2\mu^*)z^2}{((x-x_1)^2+y^2+z^2)^{5/2}} - \frac{1-2\mu^*}{((x-x_1)^2+y^2+z^2)^{3/2}}  + \\ \frac{3\mu^*z^2}{((x-x_{21})^2+y^2+z^2)^{5/2}}- \frac{\mu^*}{((x-x_{21})^2+y^2+z^2)^{3/2}}  + \\ \frac{3\mu^*z^2}{((x-x_{22})^2+y^2+z^2)^{5/2}} - 
\frac{\mu^*}{((x-x_{22})^2+y^2+z^2)^{3/2}}\bigg. \bigg],
\label{ZZ}
\end{multline}

\begin{multline}
\Omega_{xy}= \Omega_{yx}= k\bigg[ \bigg.\frac{3(1-2\mu^*)(x-x_1)^2y}{((x-x_1)^2+y^2)^{5/2}}+ \frac{3\mu^*(x-x_{21})^2)y}{((x-x_{21})^2+y^2)^{5/2}} +\\ \frac{3\mu^*(x-x_{22})y}{((x-x_{22})^2+y^2)^{5/2}}\bigg. \bigg],
\label{XYY}
\end{multline}

\begin{multline}
\Omega_{xz}= \Omega_{zx}= k\bigg[ \bigg.\frac{3(1-2\mu^*)(x-x_1)z}{((x-x_1)^2+y^2+z^2)^{5/2}}+ \frac{3\mu^*(x-x_{21})z}{((x-x_{21})^2+y^2+z^2)^{5/2}} +\\ \frac{3\mu^*(x-x_{22})z}{((x-x_{22})^2+y^2+z^2)^{5/2}}\bigg. \bigg],
\label{XZZ}
\end{multline}

\begin{multline}
\Omega_{yz}= \Omega_{zy}= k\bigg[ \bigg.\frac{3(1-2\mu^*)yz}{((x-x_1)^2+y^2+z^2)^{5/2}}+ \frac{3\mu^*yz}{((x-x_{21})^2+y^2+z^2)^{5/2}} +\\ \frac{3\mu^*yz}{((x-x_{22})^2+y^2+z^2)^{5/2}}\bigg. \bigg].
\label{YZ}
\end{multline}

In Equations \ref{AEPxx} - \ref{AEPzz}, $ \xi $, $ \eta $ and $ \zeta $ represent the position of the particle with respect to the equilibrium point. Through numerical analysis, we observed that the equilibrium points exist only in the $xy$ plane, regardless of the values assigned to $d$, $\mu^*$ and $k$.
Due to the fact that the equilibrium points for the rotating mass dipole model are in the $ xy $ plane, the Equation \ref{AEPzz} is decoupled (it does not depend on $ \xi $ and $ \eta $), therefore,  the equation of motion \ref{AEPzz} becomes
\begin{equation}
\ddot{\zeta} = -\vartheta\zeta
\label{zeta}
\end{equation}
where $\vartheta$ is constant and depends on the values assigned to $d$, $\mu^*$ and $k$. Equation \ref{zeta} shows that the motion perpendicular to the $ xy $ plane is periodic with frequency $ \omega $ = $ \sqrt{\vartheta} $. Motion in the $ z $ direction is therefore limited with
\begin{equation}
\zeta = c_3\cos(\sqrt{\vartheta}t) + c_4\sin(\sqrt{\vartheta})t
\label{zeta1}
\end{equation}
where $ c_3 $ and $ c_4 $ are integration constants.

When the motion is in the $ xy $ plane, the non-trivial characteristic roots of the Equation \ref{AEPxx}, \ref{AEPyy} were obtained in \citet{2017ApSS.362...61B} (considering $k$ = 1.). The linearization around $ L_1 $ and $ L_2 $ provides a pair of real eigenvalues ($ saddle $), corresponding to one-dimensional stable and unstable manifolds, and one pairs of imaginary eigenvalues, suggesting a two-dimensional central subspace in plane $xy$, which accounts for an oscillatory behavior around the equilibrium point of the linear system \citep{Howell1, Haapala}. Hence, in general, for $ L_1 $ and $ L_2 $, the stability type is $ saddle \times center \times center $ for the problem studied here and also for the CRTPB considering 0 $<$ $\mu$ $\leq$ 0.5. The Lyapunov Center Theorem  guarantees, for the planar case, the existence of a one-parameter family of periodic orbits emanating from each of the collinear equilibrium points. Thus, for the spatial case, two one-parameter families of periodic orbits around $L_1$ and $L_2$ are expected. 
It was observed that the nature of the eigenvalues of the collinear equilibrium points is not altered when we vary $d$, $\mu^*$ and $k$.

Consider the linearized dynamics around the $L_1$ equilibrium point. We will adopt the coordinates $\mathbf{x'}$ = ($\xi$; $\eta$; $u$; $v$), where $u$ and $v$ are the velocities in the $x$ and $y$ direction, respectively, for the physical variables in the linearized planar system.
To differentiate, we will use the coordinates $\mathbf{x_0}$ = ($x$; $y$; $\dot{x}$; $\dot{y}$) for the physical variables in the nonlinear system and, finally, $\mathbf{y}$ = ($y_1$; $y_2$; $y_3$; $y_4$) for the variables in the diagonalized system.
We know that if we choose an initial condition anywhere near the equilibrium point, the real components of the eigenvalues (stable and unstable) will dominate the particle's behavior.
But instead of specifying any initial condition for the system, we want to find an orbit around the equilibrium point $L_1$, for example, with some desired behavior, such as a periodic orbit. This becomes easy if we use the diagonalized system ($\mathbf{y_0}$) to determine the initial conditions.
As we want to minimize the component in the unstable direction of the non-linear path, we must choose the initial conditions that correspond to the harmonic motion of the linear system. Thus, we choose the initial condition in the diagonalized system as $\mathbf{y_0}$ = (0; 0; $y_3$;$y_4$), where the non-zero initial values can be complex numbers and is intended to amplify the oscillatory terms. Null terms have the function of nullifying exponential (unstable) terms. In fact, if we want to get real solutions at the $\mathbf{x'}$ coordinates, we must consider $y_3$ and $y_4$ as complex conjugates. Transforming these conditions back to the original coordinates of the linear system, from the transformation $\mathbf{x_0}$ = $T\mathbf{y_0}$, we find the initial conditions in the linearized system $\mathbf{x'_0}$ = ($x'$, $y'$, $u$, $v$), where $T$ is the matrix of the eigenvectors
of the state transition matrix $A$. The Jacobian matrix $A$ contains the pseudo potential Hessian, derived from the truncated Taylor series expansion over the reference solution. 

Due to the fact that these initial conditions were chosen to nullify the unstable and stable eigenvectors, they provide a harmonic movement in the linear system.

Now that we have the initial conditions for the linear system, we want to find a periodic planar orbit in the nonlinear system.

We can note that the potential function for the system studied here depends only on the distances that a spacecraft are from the primary bodies, that is, it has symmetry with respect to the $x$-axis. Taking advantage of the fact that the planar orbits are symmetrical with respect to the $x$-axis, the initial state vector takes the form $\mathbf{x_0}$= [$x_0$ 0, 0, $\dot{y}_0$]$^T$.
These symmetries were used to find symmetric periodic orbits. This is done by determining the initial conditions, on the $x$-axis, where the initial velocity is perpendicular to this axis ($\dot{y}$) and then the integration is done until the path returns by crossing the $x$-axis with the speed orientation $\dot{y}_f$ opposite to the initial condition. This orbit can be used as an initial guess to use
Newton's method, where the target state is quoted above; that is, that the orbit returns to $x$-axis with normal velocity. The equations of motion and the State Transition Matrix are incorporated
numerically until the trajectory crosses the $x$-axis again. The final desired condition has the following form
$\mathbf{x_f}$= [$x_f$ 0, 0, $\dot{y}_f$]$^T$.

\section{Collinear equilibrium points as a function of the ratio between gravitational and centrifugal accelerations}
\label{CEP}

In this section, we analyze the influence of the parameter $k$ on the position of the collinear equilibrium points, since the influence of $d$ and $\mu^*$ on the collinear points has already been performed in the work of \citep{2017ApSS.362...61B}.

To determine how $ k $ affects the positions of the collinear equilibrium points, we consider $ \mu^* ~ = ~ 1 \times10^{-3} $ and $ d ~ = ~ 1/12 $ canonical units.

\begin{figure}[!htbp]
	\centering\includegraphics[scale=0.5]{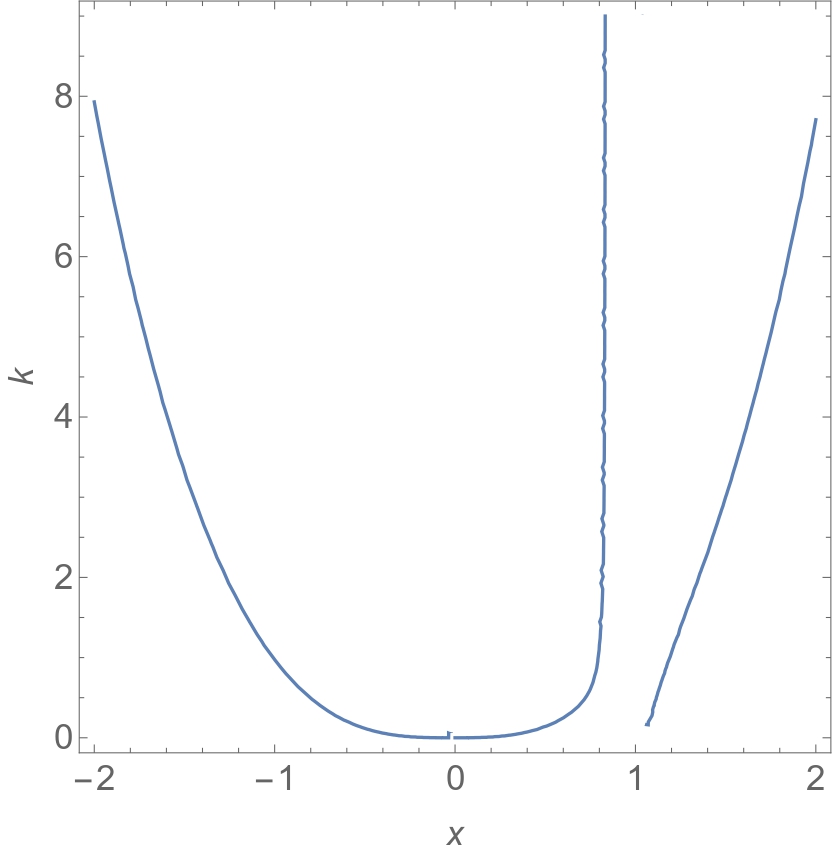}
	\caption{$x$-coordinates of the equilibrium points $L_1$, $L_2$ and $L_3$ for different values of $k$.}
	\label{kchange}
\end{figure}

Figure \ref{kchange} shows the $ x $ coordinates of $ L_1 $, $ L_2 $ and $ L_3 $ as a function of $ k $. Because they are at both ends of the $ x $ axis, the positions of $ L_2 $ (right curve) and $ L_3 $ (left curve) are more affected than the position of $L_1$. Consider that there are three forces acting on the system: (i) the gravitational force of $ M_1 $; (ii) the gravitational attraction of $ M_2 $; and (iii) the centrifugal force, which is directly proportional to the angular velocity of the system around the center of mass and the distance between the equilibrium point and the center of mass of the system. Thus, by decreasing the angular velocity of the asteroid system around the center of mass, as $ k $ becomes larger, it is necessary to increase the distance between $P$ and the center of mass such that the centrifugal force remains at the same value and it counterbalances the gravitational forces from $ M_1 $ and $ M_2 $, which remain unchanged. Thus,  $ L_2 $ and $ L_3 $ move away from the center of mass of the system. Although $ L_1 $ also moves away from the center of mass of the system, it does so in a more subtle way. This is because, when moving away from the center of mass of the system, $ L_1 $ approaches $ M_2 $. Regarding the gravitational force increases, a balancing force is needed to prevent $ L_1 $ from going too close to $ M_2 $.

As shown in Figure \ref{kchange}, the $x$ coordinates of $ L_2 $ and $ L_3 $ tend to $\pm ~\infty$, respectively, when $ k ~ \rightarrow ~ \infty $, that is, when the asteroid system ceases to rotate. This implies that $ L_2 $ and $ L_3 $ cease to exist when the asteroids are static. On the other hand, the equilibrium point $ L_1 $ continues to exist when $ k ~ \rightarrow ~ \infty $, due to the balance between the gravitational forces between $ M_1 $ and $ M_2 $.

\section{Periodic orbits around the first and second collinear equilibrium points as a function of the mass parameter and the size of the dipole}
\label{POAFSC}

Based on previous knowledge about Type A asteroids, we consider that the most massive primary is spherical in shape and with a diameter of 5 km \citep{2007Icar..190..250P, 2015aste.book..375W}. Also, knowing that, on average, the mutual orbit of type A binary asteroids has a semi-major axis of about 4.8 primary component radii \citep{2015aste.book..375W}, we consider that the distance between the bodies is 12 km, which is the normalization factor for the distances. Finally, type A asteroids are known to have moderately sized secondary, ranging from 4\% to 58\% of the size of the primary, whose mass ratio ($m_2/(m_1+m_2$)) range varies from $6.4\times10^{-5}$ to $2.0\times10^{-1}$. Based on this evidence, we will consider in this analysis the dimension of the secondary body from 0 to 2 km, where we vary in step $500$ meters, and a range of the mass ratio from 1$\times 10^{-5}$ to 1$\times 10^{-1}$, where we vary in step $10^{-1}$.

Periodic orbits are of special interest to explore the dynamical behavior of a massless particle in the vicinity of two primary bodies.

The results below were obtained by calculating approximately 3500 orbits from each family, starting from an initial condition with very low amplitude, and continuing the families until the orbits obtained came near the surface of the asteroids. To find symmetric periodic orbits, we consider $ k = 1 $, that is, the bodies orbit the center of mass of the system without any internal forces.

Each family was calculated for different values of $ \mu^* $ and $ d $ to highlight the effect of the mass ratio of the system and of the elongated shape of the secondary body on the dynamical behavior of a space vehicle in the vicinity of the binary system. To analyze the influence of the elongation of the secondary body on the periodic orbits, we determined the periodic orbits considering the values $d$ = 0; 0.5; 1; 1.5 and 2 km. Also aiming to understand the influence of $\mu^*$ on the periodic orbits, we determine the periodic orbits considering the values $\mu^*$ = $10^{-5}$; $10^{-4}$; $10^{-3}$; $10^{-2}$ and $10^{-1}$.

We are interested in the stability of the periodic solutions, which can be determined by analyzing the eigenvalues of  the monodromy matrix. 
Given the sympletic nature of the dynamical system, if $\lambda$ is a characteristic multiplier, then $1/\lambda$ is also, as well as,  $\overline{\lambda}$  and $1/\overline{\lambda}$. Thus, the periodic solutions investigated have six characteristic multipliers that appear in reciprocal pairs, with two of them being unitary \citep{Kenneth, natasha2}. 
The other four may be associated with the central subspace or with  the stable/unstable subspace. In general, a particular orbit has six characteristic multipliers of the form $ 1 $, $ 1 $, $ \lambda_1 $, $ 1/\lambda_1 $, $ \lambda_2 $ and $ 1/\lambda_2 $.

The stability indices offer a useful measure of orbital stability. Following \citep{1969AIAAJ...7.1003B}, the stability index is defined as $ s_i $ = $ |\lambda_i  $ + $ 1 /  \lambda_i | $, $ i = 1, 2$. A periodic orbit is unstable and there is a natural flow out and into the orbit if any stability index is greater than 2, that is, if $ s_i $ $> $ 2. On the other hand, a periodic orbit is stable and has no unstable subspace, that is, if $ s_i $ $ <$ 2 \citep{2020CeMDA.132...28Z}.
The magnitude of the stability index is directly related to the arrival/departure flow rate. The higher the value of $ s_i $, the more unstable is the periodic orbit and bifurcations can occur when $ s_i = 2$.

Given that the periodic orbits growing from the collinear points inherit the stability properties of $L_1$, $L_2$, and $L_3$, 
the eigenvalues of the monodromy matrix of these orbits and corresponding stability indices appear as:
(i) a trivial pair of unitary values, resulting in $s_0$ = 2; 
(ii) a real pair of reciprocals , resulting in $s_1>$ 2; and 
(iii) a pair of complex conjugate eigenvalues with unitary absolute value, implying $s_2$ $<$ 2. Thus, given that, for the subsets of the periodic orbit (PO) families near the equilibria, $s_1$ is related to the stable/unstable subspace ($ \lambda^{W_s} $/ $ \lambda^{W_u} $), while $s_2$ is the stability index corresponding to the pair accounting for the central subspace. 

\subsection{Planar orbits}

\begin{figure}[!htbp]
	\centering\includegraphics[scale=0.55]{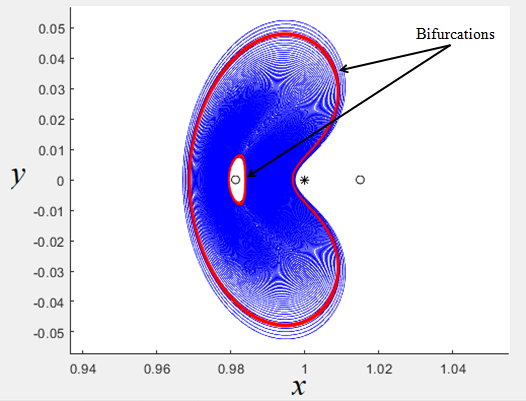}
	\caption{Planar orbits around of the equilibrium point $L_1$ considering $\mu^* = 10^{-5}$ and $d$ = 0.}
	\label{planar3}
\end{figure}

Figure~\ref{planar3} shows a family of planar orbits around $ L_1 $ with $ \mu^* = 10^{-5} $ and $ d $ = 0.  The orbits obtained do not intersect the asteroid although, as seen in Figure \ref{planar3}, as the amplitude increases along the family, the orbits expand from the vicinity of the equilibrium point towards the surface of the secondary body (black asterisk). 

In Figure \ref{planar3}, the red orbits indicate where bifurcations occur, that is, when one of the stability indices $s_1$ or $s_2$ reaches the critical value 2. Note in Figure \ref{planar3} that the maximum position, when the second bifurcation occurs, reached by the infinitesimal mass body in the $x$ component is greater than the position of the secondary body.

Although many bifurcations exist in dynamical systems, only two types of bifurcation are of particular interest for the focus of this work; the pitchfork and period-multiplying bifurcations.

A family of periodic orbits undergoes a pitchfork bifurcation when the stability of the periodic orbit changes as a parameter evolves, which in our case is the energy constant. During this type of local bifurcation, a pair of eigenvalues (not trivial) of the monodromy matrix pass through the critical values $ \lambda_1 =  1/\lambda_1$ (or $\lambda_2 =  1/\lambda_2$) = + 1 of the unit circle. Consequently, the stability index passes through $s_1$ (or $s_2$) = 2 \citep{natasha2}. In addition, the stability of the periodic orbits changes along a family, an additional family of a similar period is formed. This new family of orbits has the same stability as the members of the original family before the bifurcation arose.
On the other hand, a period-doubling bifurcation is identified when a pair of not trivial eigenvalues ($ \lambda_{1,2} $ and $ 1/\lambda_{1,2} $, where $\lambda_{1,2}$ means $\lambda_1$ or $\lambda_2$),  passes through $ \lambda_{1,2} = 1/\lambda_{1,2} $  = - 1 of the unit circle. Therefore, it represents a critical value of the stability index, such that $s_{1,2}$ = - 2  \citep{natasha2}.

\begin{figure}[!htbp]
	\centering\includegraphics[scale=0.52]{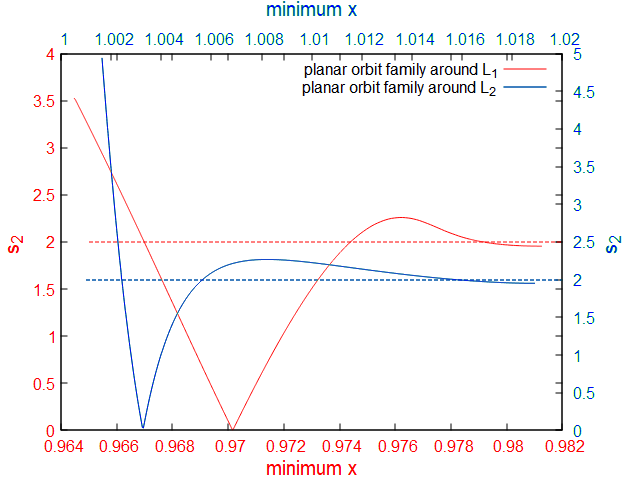}
	\caption{Stability index ($ s_2 $) around $ L_1 $ (red) and $L_2$ (green) considering $ d ~ = ~ 0 $ and $\mu^* = 10^{-5}$.}
	\label{index}
\end{figure}

When building the families of planar orbits, with $ d ~ = ~ 0 $ and $ \mu^* $ $ = $ $ 10^{-5} $, we observe that the stability index ($ s_2 $) reaches the critical value three times for planar orbits around $ L_1 $ and $ L_2 $, as seen in Figures~\ref{index}. In both figures, the horizontal axis display the minimum $x$ value along the orbits. The stability index $ s_1 $ does not reach the critical value for this $ \mu^* $ and $ d $.

For $ \mu^* = 10^{-5} $, the equilibrium point $ L_1 $ is located at $ x ~ = ~ 0.981278 $, $ y ~ = ~ 0 $ and $ z ~ = ~ 0 $, while $ L_2 $ is at position $ x ~ = 1.01892 $, $ y ~ = ~ 0 $ and $ z ~ = ~ 0 $. The orbits with smaller amplitudes are close to the equilibrium point (right side of Figures \ref{index}) and the first bifurcation occurs for a small amplitude orbit ($ x ~ \approx $ 0.97583 for $ L_1 $ and $ x ~ \approx $ 1.01577 for $ L_2 $). As we continue the planar family, the stability index $s_2$ shown in Figure \ref{index} continues to increase, reaches a maximum, decreases and reaches the value 2 again, where another bifurcation occurs, with $ x ~ \approx $ 0.99408 for $ L_1 $ and $ x ~ \approx $ 1.0056 for $ L_2 $. 
As we continue the families of planar orbits around $ L_1 $ and $ L_2 $, the stability index decreases, reache a minimum, increases and again and reaches the critical value 2, where another bifurcation occurs. After the third bifurcation, the stability index further increases and we did not detect additional bifurcations given that, as the orbits are very close to the center of mass of the secondary body, our Newton method looses track of planar orbits, converging to a completely different family of orbits.

\begin{figure}[!htbp]
	\centering\includegraphics[scale=0.5]{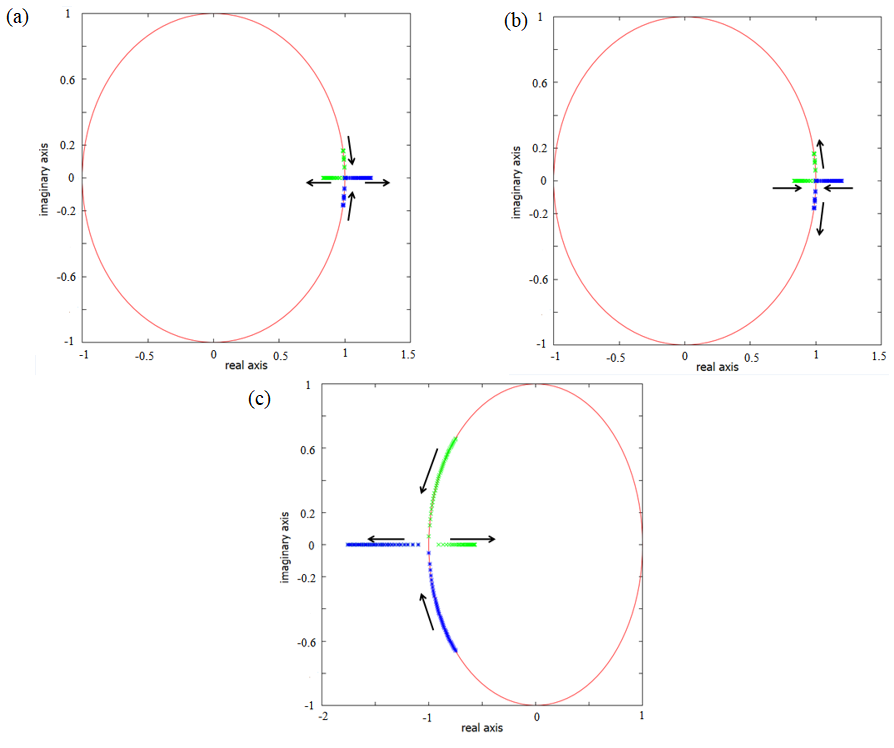}
	\caption{(a) Behavior of the characteristic multipliers at the first pithckfork bifurcation around $L_1$ and $L_2$. (b) Behavior of the characteristic multipliers at the second pithckfork bifurcation around $L_1$ and $L_2$. (c) Behavior of the characteristic multipliers that leads to the period-doubling bifurcation around $L_1$ and $L_2$. In these cases we consider $d ~=~0$ and $\mu^*$ $=$ $10^{-5}$.}
	\label{1}
\end{figure}

Figures \ref{1} (a), (b) and (c) provide information about the types of the bifurcations that occur along the family of planar orbits. For $ \mu^* = 10^{-5} $ and $ d = $ 0, analyzing the path of the characteristic multipliers in Figures \ref{1} (a) and (b), we find that the first bifurcation is a supercritical pitchfork bifurcation, while the second one corresponds to a subcritical pitchfork  case. This suggests that new families of periodic orbits appear in those regions when the bifurcation occurs \citep{2016AdSpR..58..387F}. In fact, after the first bifurcation (low amplitude periodic orbit), it is possible to detect halo orbits, while after the second bifurcation the family of axial orbits appears \citep{Grebow}. Unlike the planar Lyapunov orbits, halo and axial orbits are three-dimensional.

Figure \ref{1} (c) shows the behavior of the eigenvalues at the third bifurcation. The characteristic multipliers start in the imaginary plane and move until they collide on the negative real axis and start to obtain only real values on the negative axis. Thus, the eigenvalues indicate a period-doubling bifurcation. 

\begin{figure}[!htbp]
	\centering\includegraphics[scale=0.7]{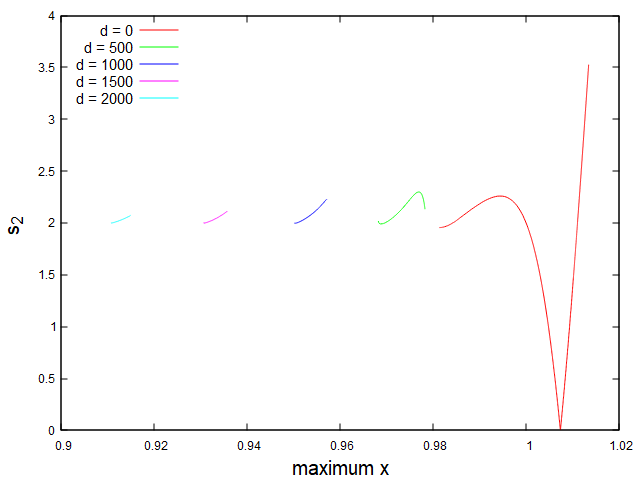}
	\caption{Planar orbit stability index around $ L_1 $ for different values of $ d $.}
	\label{dchange}
\end{figure}

\begin{figure}[!htbp]
	\centering\includegraphics[scale=0.7]{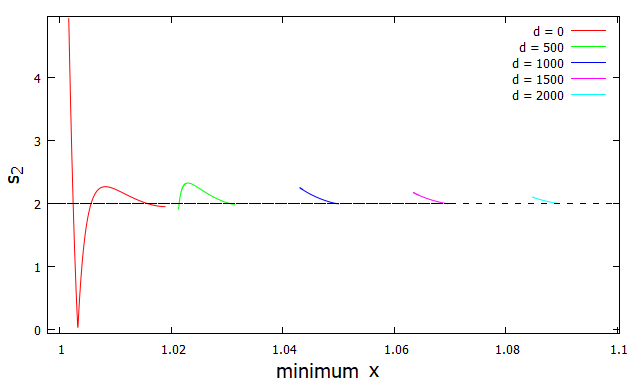}
	\caption{Planar orbit stability index around $ L_2 $ for different values of $ d $.}
	\label{dchangel2}
\end{figure}

Figure \ref{dchange} and \ref{dchangel2} provides information about the stability index (considering  the values of $ s_2 $), around $L_1$ and $L_2$, respectively, when we increase the dipole dimension from 0 meters, that is, the body is modeled as a mass point, up to the dimension of 2000 meters. In this analysis we consider the constant mass ratio in the value of $ \mu^* ~ = ~ 10^{-5} $. When we consider the dipole as a point mass body ($ d $ = 0), it is possible to observe three bifurcations (the red curve passes through the critical value three times). We can observe that as we increase the dimension of the secondary, the second bifurcation points in the planar orbits around $ L_1 $ and $L_2$ cease to exist because the trajectories collide with the secondary body.
This is because, as the dimension of the dipole varies and the planar orbits approach the secondary body, our Newton method looses track of the planar orbits, converging to a completely different family of the orbits.
Note that, the larger the dipole size, the smaller the planar orbit family found.

\subsection{Influence of the mass parameter and the size of the dipole on the planar orbits}

Now, we investigate how the planar orbits evolve as a function of the dipole size and mass ratio in canonical units. With the normalization factor being $ D =  12000 $ meters, the dipoles sizes used in our study were $ d $ = 0, 500, 1000, 1500 and 2000 meters.

\begin{figure}[!htbp]
	\centering\includegraphics[scale=0.35]{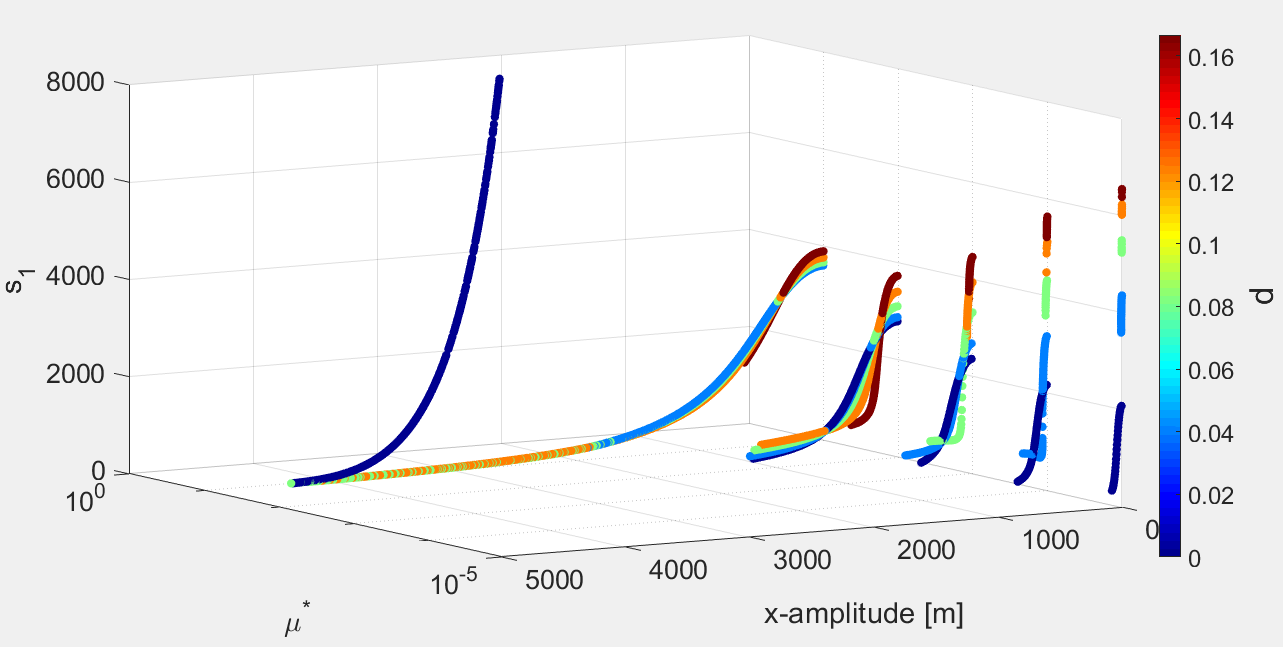}
	\caption{Stability index ($ s_1 $) of the planar orbits around $ L_1 $ for different values of $ d $ and $ \mu^* $.}
	\label{SI(T)Lyapunov}
\end{figure}

Figures \ref{SI(T)Lyapunov} provide information about the stability index $ s_1 $ of the planar orbits  around $ L_1 $, respectively, as a function of $ d $ and $ \mu^* $. In both figures, the color code accounts for the size of the dipole ($ d $). 

First, we investigate the solutions as $ d $ varies and $ \mu^* $ is kept constant. Note that, in Figures \ref{SI(T)Lyapunov}, in general, when the size of the dipole increases, the planar orbits become more unstable. This means that, the larger the secondary body, the more unstable the planar orbits are.

If we consider $ d $ = 0, which corresponds to the CRTBP, we observe that, as $ \mu^* $ increases, the orbits become increasingly unstable. On the other hand, when the elongated form of the secondary body is taken into account, $ s_1 $ becomes smaller as $ \mu^* $ increases, and it only increases again after $ \mu^* $ = $ 10^{-1} $. This information is important for space missions, since a high value in the stability index ($ s_i $) indicates a divergent mode that moves the spacecraft away from the vicinity of the orbit quickly. In general, the stability index is directly related to the space vehicle's orbital maintenance costs and inversely related to the transfer costs. This same analysis was performed around the $L_2$ equilibrium point, where we found similar results.

\begin{figure}[!htbp]
	\centering\includegraphics[scale=0.35]{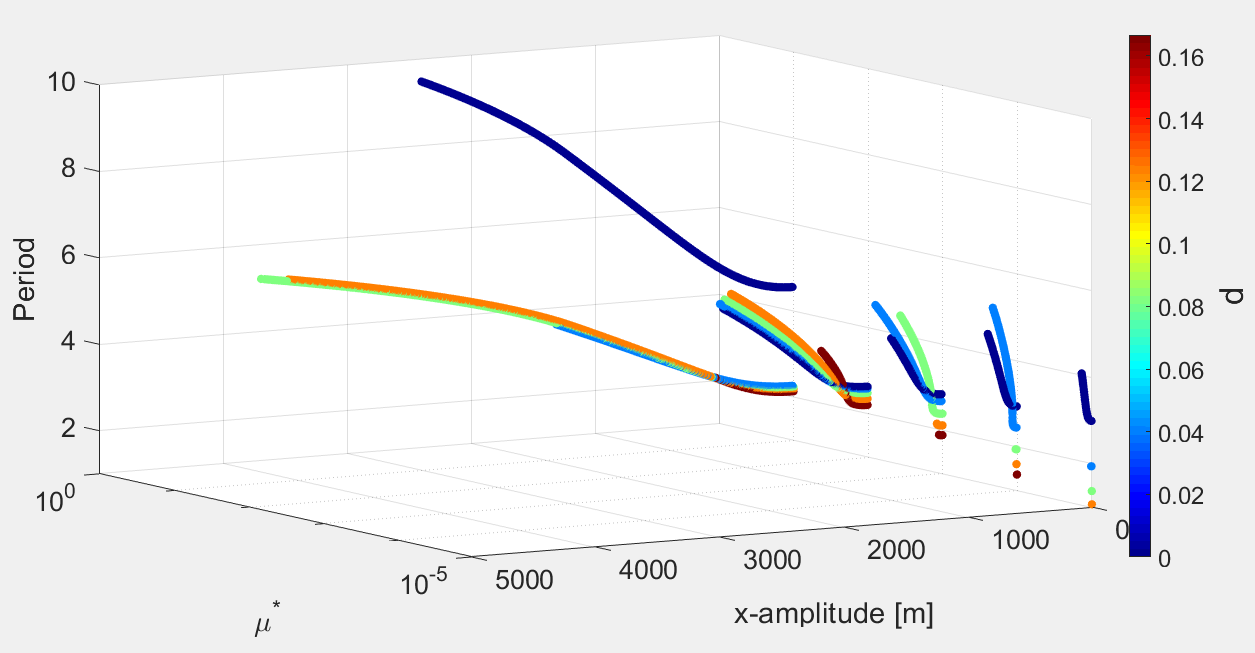}
	\caption{Period of planar orbits around $L_1$ for different values of $ d $ and $ \mu^*$.}
	\label{Period(top)Lyapunov}
\end{figure}

Next, we analyze the period of the planar orbits in terms of $ d $ and $ \mu^* $ around $L_1$. As shown in Figures \ref{Period(top)Lyapunov}, for low amplitude, as $ d $ increases, with $ \mu^* $ kept constant, the period of the planar orbits decreases.
This is because the mass distribution of the secondary body allows part of the mass of the asteroid to be closer to the negligible mass particle, causing the gravitational attraction to become larger, thus increasing the acceleration in the vicinity of the secondary body and decreasing the orbital period. On the other hand, when the $x$-amplitude is large, the results can be inverted, as shown in \ref{Period(top)Lyapunov}. In general, when the amplitude of the orbit increases, the orbital period becomes longer. Similar results were found in the vicinity of $L_2$.


Considering the family with $ d = 0 $, when $ \mu^* $ increases, the period of the orbits remains similar, except when $ \mu^* $ = $ 10^{-1} $. Conversely, when the elongation of the secondary body is considered, in general, for a given value of $ d $, the larger the mass ratio, the longer the orbital period.

\begin{figure}[!htbp]
	\centering\includegraphics[scale=0.33]{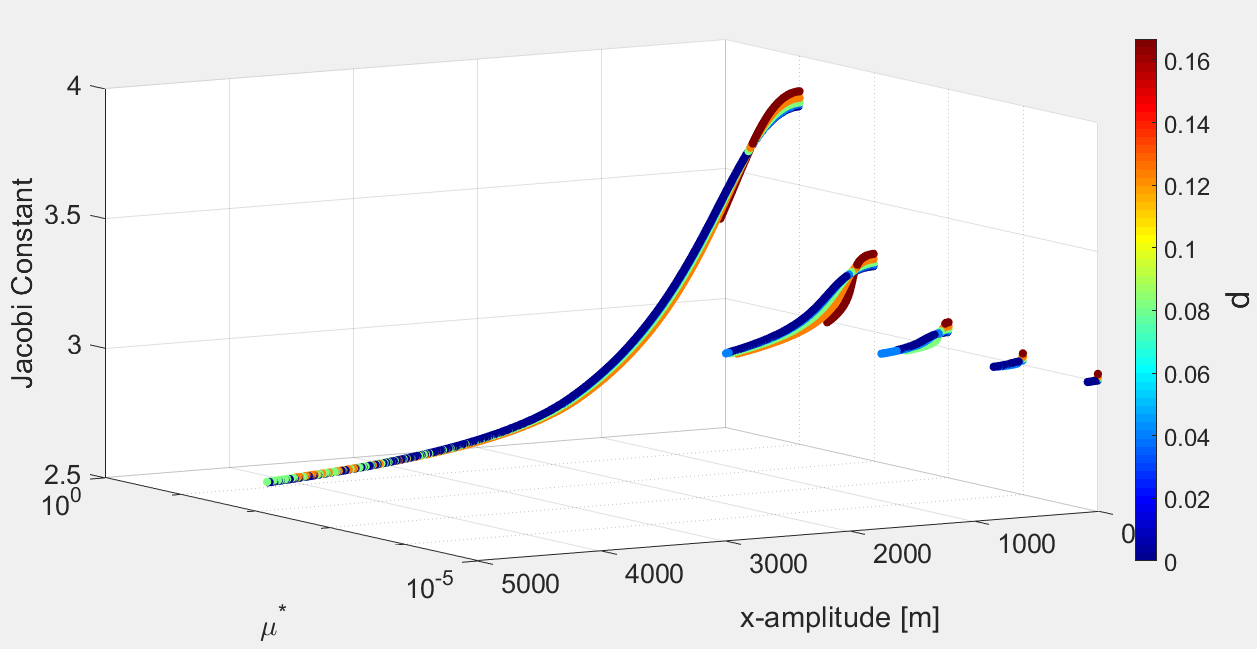}
	\caption{Jacobi constant of planar orbits around $L_1$ for different values of $ d $ and $ \mu^* $.}
	\label{Jacobi_constant_Lyapunov_L1}
\end{figure}

Finally, we analyze the energy of the system in terms of $ d $ and $ \mu^* $, as shown in the Figures \ref{Jacobi_constant_Lyapunov_L1}.

We find that when $ d $ or $ \mu^* $ increases, the energy required to orbit a given equilibrium point decreases. That is, the more elongated the secondary body and the larger the value of $\mu^*$, the less energy is needed to orbit a given equilibrium point. This also means that, as the size of the dipole increases or as the mass ratio of the system increases, the bifurcations occur at lower energies. The same analysis performed for $L_1$ can be done for $L_2$.

\subsection{Computing halo orbits}

\begin{figure}[!htbp]
	\centering\includegraphics[scale=0.65]{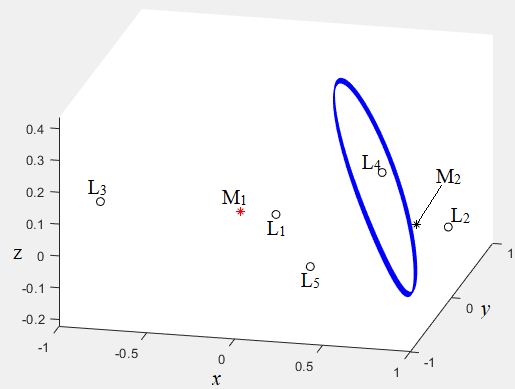}
	\caption{Halo orbits around the $L_1$ equilibrium point considering $\mu^* = 10^{-5}$ and $d$ = 0.}
	\label{halo_orbit_around}
\end{figure}

Halo orbits are a three-dimensional branch of planar orbits that appear when the planar orbit stability index reaches the critical value $ s_2 = 2 $. Figure \ref{halo_orbit_around} shows a family of halo orbits around $L_1$ with $\mu^*$ = $10^{-5}$ and $d$ = 0. The orbits are in three-dimensional space and as the amplitude increases along the family, the halo orbits expand from the vicinity of the equilibrium point towards the surface of the secondary (black asterisk). 

To find the initial conditions of the halo orbit, we keep the coordinate $x_0$ fixed and search for $z_0^*$, $\dot{y}_0^*$ and $T/2^*$ such that $\dot{ x}^*(T/2^*)$, $\dot{z}^*(T/2^*)$ and $y^*(T/2^*)$ are all null. Then, to find the halo orbit, we use as initial guess the position $x_0$, velocity ($\dot{y}$) and period ($T$) of the planar orbit when the stability index $s_2 = 2$. Knowing these initial conditions, all that remains is to determine the initial guess of the position on the $z$ axis, such that we can find the halo orbit. Because the halo and planar orbit are similar (when $s_2 = 2$), the position on the $z$ axis of the halo orbit must have a very small value (almost planar orbit). Thus, in this work, the value of $z_0$ = 0.0001 canonical unit was used as the initial guess for the position on the axis $z$.
A Newton method for this problem is
\begin{equation}
\mathbf{x}_{n+1} = \mathbf{x}_n - [Df(\mathbf{x}_n)]^{-1}f(\mathbf{x}_n)
\label{newtonme}
\end{equation}
with $\mathbf{x}$ = ($z, \dot{y}$, $T/2$) and $\mathbf{x}_0$ = ($z_0, \dot{y}_0$, $T_0/2$). Aqui ($z_0$, $\dot{y}_0$, $T_0/2$) is the initial guess of the halo orbit.

The differential is
\begin{equation}
Df(\mathbf{x} = 
\begin{bmatrix}
\phi_{4,3} & \phi_{4,5} &  g_4(x_0,0,z(T/2),0,\dot{y}(T/2),0) \\ 
\phi_{6,3} & \phi_{6,5} & g_6(x_0,0,z(T/2),0,\dot{y}(T/2),0)\\ 
\phi_{2,3}& \phi_{2,5} &g_2(x_0,0,z(T/2),0,\dot{y}(T/2),0)
\end{bmatrix}
\label{df}
\end{equation}
where $\phi_{i,j}$ are elements of the monodromy matrix, $g: U \subset \mathbb{R}^6 \rightarrow \mathbb{R}^6$ is the vector field of the restricted synchronous three-body problem, $z(T/2) = \phi_3$($x_0, 0, z, 0, y, 0, T/2$) and $\dot{y}(T/2)$ = $\phi_5$($x_0, 0, z, 0, y, 0, T/$). With this information, we expect that if $\mathbf{x}_0$ is close enough to the halo orbit, then $\mathbf{x}_n$ $\rightarrow$ $\mathbf{x}^*$ as $ n \rightarrow \infty$. $g$ is given by Equation \ref{g}.
\begin{multline}
g(x,y,z,\dot{x}, \dot{y}, \dot{z}) = 
\begin{bmatrix}
g_1(x,y,z,\dot{x}, \dot{y}, \dot{z})\\ 
g_2(x,y,z,\dot{x}, \dot{y}, \dot{z})\\ 
g_3(x,y,z,\dot{x}, \dot{y}, \dot{z})\\
g_4(x,y,z,\dot{x}, \dot{y}, \dot{z})\\
g_5(x,y,z,\dot{x}, \dot{y}, \dot{z})\\
g_6(x,y,z,\dot{x}, \dot{y}, \dot{z})
\end{bmatrix}
=
\begin{bmatrix}
\dot{x}\\ 
\dot{y}\\ 
\dot{z}\\ 
2\dot{y} + D_x\Omega\\ 
-2\dot{x} + D_y\Omega\\ 
D_z\Omega
\end{bmatrix}
\label{g}
\end{multline}
All the information we need to start Newton's method is shown above.

From the cylinder theorem, it was possible to find a halo orbit family. Thus, having found a halo orbit and noticing that it has exactly two unit eigenvalues, we can use that as a starting point to move along the cylinder. We use the initial conditions from the previous halo orbit as a starting point to find the next halo orbit at a slightly larger value of $x$ ($x$ coordinate closer to the secondary asteroid). If we find another halo orbit here, we iterate through the process. In this way it was possible to calculate a halo orbit family. The $x$ coordinate step to determine each halo orbit was $x$ = 0.00002.

\subsection{Halo orbits}
\begin{figure}[!htbp]
	\centering\includegraphics[scale=0.65]{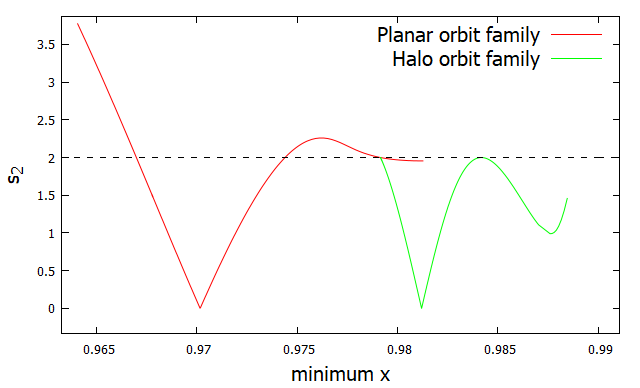}
	\caption{Stability index ($s_2$) of the planar and halo orbit families around $L_1$ considering $d = 0$ and $\mu^*$ = $10^{-5}$.}
	\label{double}
\end{figure}

\begin{figure}[!htbp]
	\centering\includegraphics[scale=0.65]{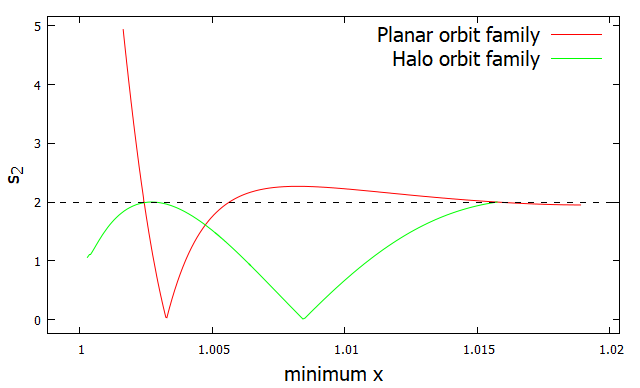}
	\caption{Stability index ($s_2$) of the planar and halo orbit families around $L_2$ considering $d = 0$ and $\mu^*$ = $10^{-5}$.}
	\label{double2}
\end{figure}

Figures \ref{double} and \ref{double2} illustrate how the halo orbits appear at the tangent bifurcations of the planar orbits around $ L_1 $ and $ L_2 $ when $ \mu^* = 10^{-5} $ and $d$ = 0. As we built the family of halo orbits, we observed that the amplitude of the orbit increases as the halo orbits move towards the secondary.

\begin{figure}[!htbp]
	\centering\includegraphics[scale=0.65]{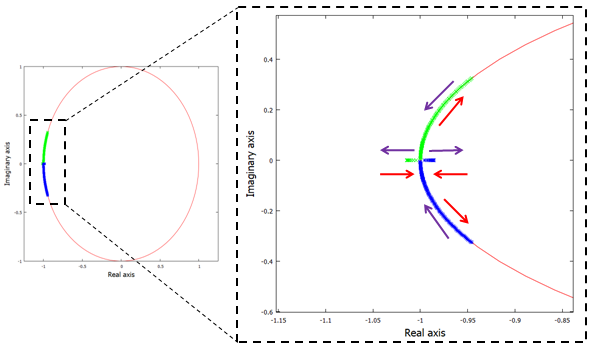}
	\caption{Behavior of the of characteristic multipliers at the period doubling bifurcation.}
	\label{doubling}
\end{figure}

For the conditions considered here, the halo orbit appears at $ x $ $ \approx $ 0.98418 for $ L_1 $ and at $ x $ $ \approx $ 1.01575 for $ L_2 $. Figure \ref{doubling} shows the path of the characteristic multipliers over the unit circle to the halo orbit around $ L_1 $ and $ L_2 $. Initially, the characteristic multipliers move in the direction shown by the purple arrows until they collide with the negative real axis, configuring a periodic doubling bifurcation. After moving subtly along the real negative axis, the characteristic multipliers return, moving in the direction of the red arrows, colliding again at -1 and then assuming imaginary values, configuring another periodic doubling bifurcation.

\begin{figure}[!htbp]
	\centering\includegraphics[scale=0.33]{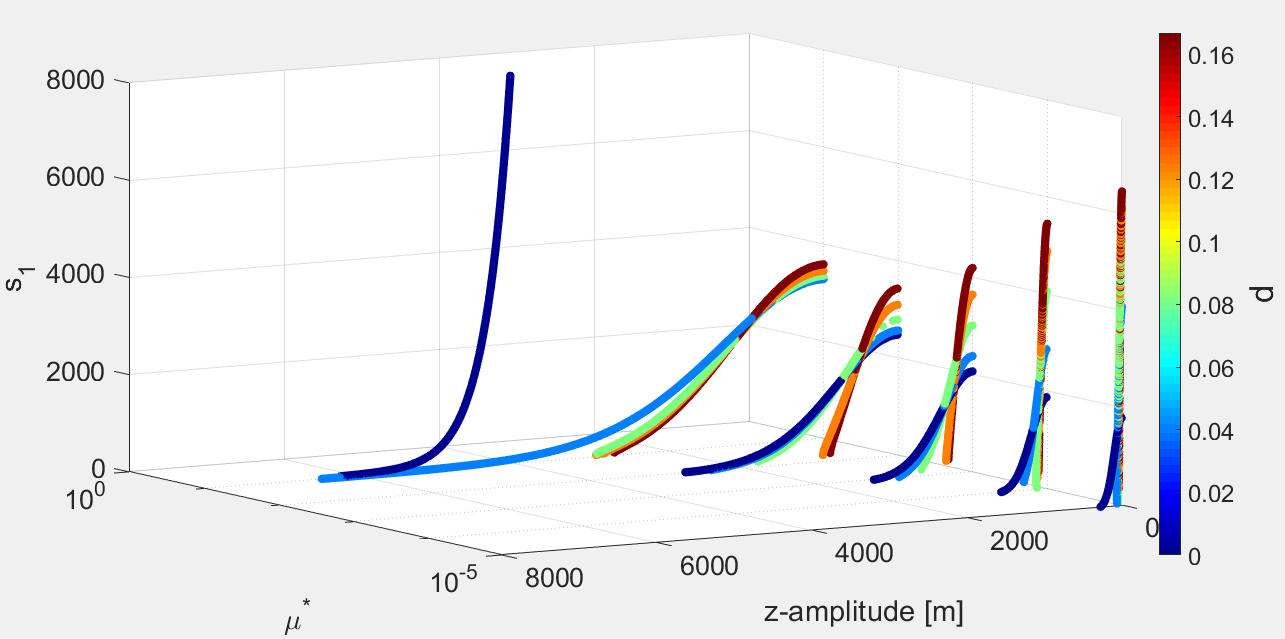}
	\caption{Stability index ($ s_1 $) of halo orbits around $ L_1 $ as a function of $ d $ and $\mu^*$.}
	\label{SI(T)halo}
\end{figure}

\begin{figure}[!htbp]
	\centering\includegraphics[scale=0.33]{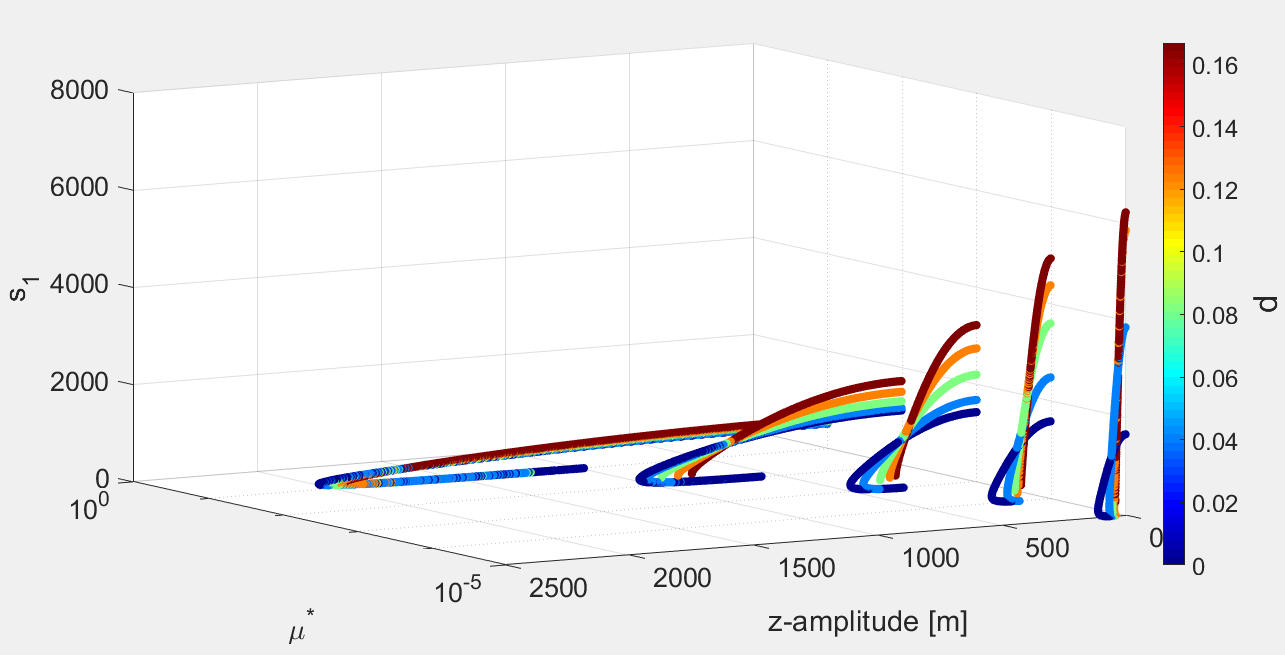}
	\caption{Stability index ($ s_1 $) of halo orbits around $ L_2 $ as a function of $ d $ and $\mu^*$.}
	\label{halol2}
\end{figure}

Figures \ref{SI(T)halo} and \ref{halol2} provide information about the $ s_1 $ stability index as a function of $ d $ and $\mu^*$. Note that the smaller the amplitudes of the halo orbits, the larger the value of the stability index $ s_1 $, when considering fixed $ d $ and $ \mu^* $. As the amplitude of the halo orbit increases, the stability index decreases. If we set $ d $ = 0, we still detect stable halo orbits for small values of $ \mu^* $. These orbits were also found by several authors using the Restricted Three-Body Problem and are called Near Rectilinear Halo Orbits (NRHO) \citep{Howell1, 2020CeMDA.132...28Z}. NRHOs, are defined as the subset of the halo orbit family with stability indexes around $ s_i $ $ \pm $ 2 and with no stability index considerably greater in magnitude than the others.

\begin{figure}[!htbp]
	\centering\includegraphics[scale=0.505]{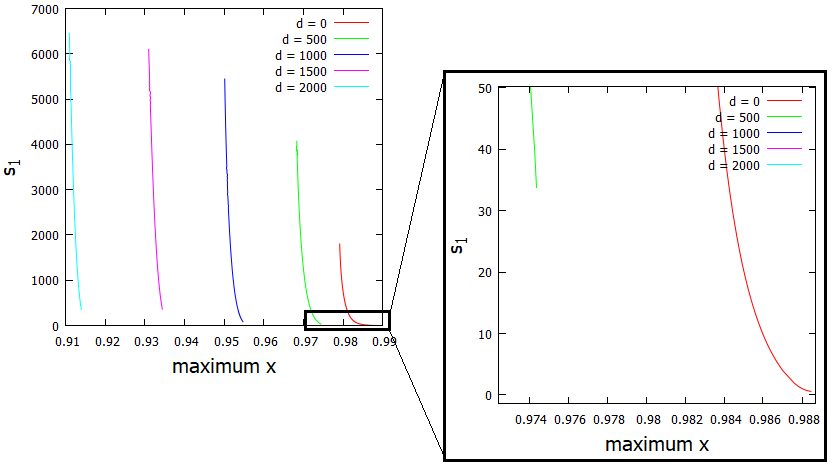}
	\caption{Halo orbit stability index around $ L_1 $ for different values of $ d $.}
	\label{s1changed}
\end{figure}

\begin{figure}[!htbp]
	\centering\includegraphics[scale=0.55]{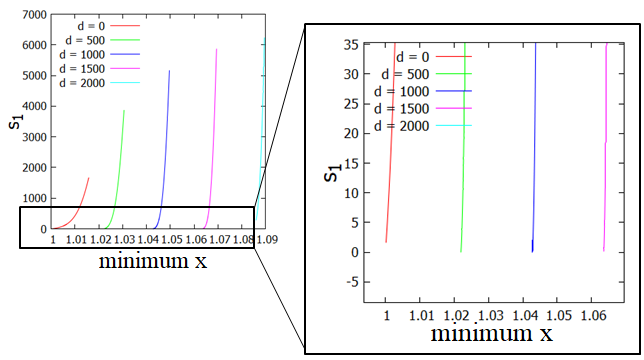}
	\caption{Halo orbit stability index around $ L_2 $ for different values of $ d $.}
	\label{s1changedl2}
\end{figure}

Figures \ref{s1changed} and \ref{s1changedl2} provide information about the stability index $ s_1 $ as the size of the dipole increases from 0 to 2000 meters and the mass ratio is kept constant at $ \mu^* $ = $ 10^{-5} $. The influence of the dimension of the secondary body on the stability of the halo orbits is clear in that plots. Note that, as the size of the secondary increases, the values of $ s_1 $ become larger in the vicinity of the equilibrium point $ L_1 $ and $ L_2 $.

Note that it is unlikely to detect NRHOs around $ L_1 $ when we take into account the elongated shape of the secondary body and assume small values of $ \mu^* $. On the other hand, there are several NRHOs around $ L_2 $. In this work, we found NRHOs up to $ d $ = 1500 meters, as shown in Figure \ref{s1changedl2}.

However, as shown in \citet{Howell1}, the stability index also depends on the mass ratio of the system. Considering $ d $ = 0 and increasing $ \mu^* $, the stability index $ s_1 $ increases. We did not detect any NRHO for values of $ \mu^* $ $ \geq $ $ 10^{-1} $ and $ d ~ = ~ 0 $. On the other hand, we find NRHO for $ \mu^* $ $ \geq $ $ 10^{-1} $ and $ d ~ = ~ 0 $ around $ L_2 $. These results are similar to those obtained by \citet{Howell1}. On the other hand, taking into account the elongation of the secondary and assuming large values of $ \mu^* $ ($ \mu^* $ $ \geq $ $ 10^{-1} $), it is possible to find family members of stable halo orbits around $L_1$ and $L_2$. Thus, in the model used in this article, stable periodic orbits in the vicinity of irregular bodies exist, even when the secondary has non-spherical shape. This agrees with the results obtained by \citet{Chappaz}, who found stable orbits around $ L_1 $ and $ L_2 $ taking into account the elongated shape of the secondary body and considering $ \mu $ = 0.4 with the triaxial ellipsoid model.

\begin{figure}[!htbp]
	\centering\includegraphics[scale=0.35]{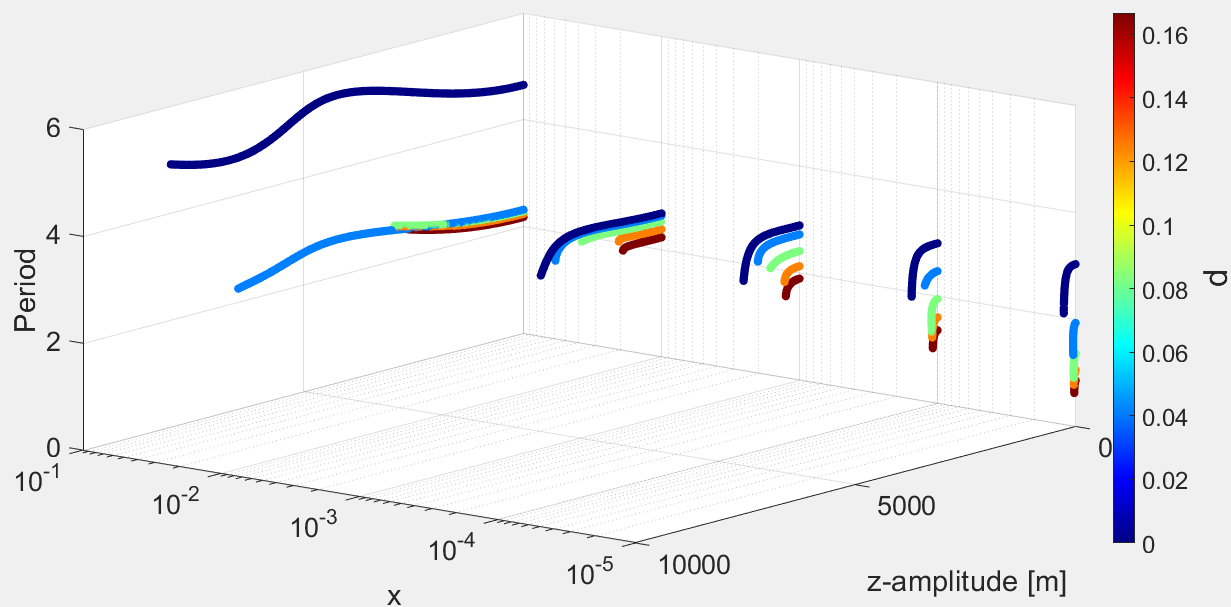}
	\caption{Period of the halo orbits around $ L_1 $ as a function of $ d $ and $ \mu^* $.}
	\label{Periodhalo}
\end{figure}

\begin{figure}[!htbp]
	\centering\includegraphics[scale=0.35]{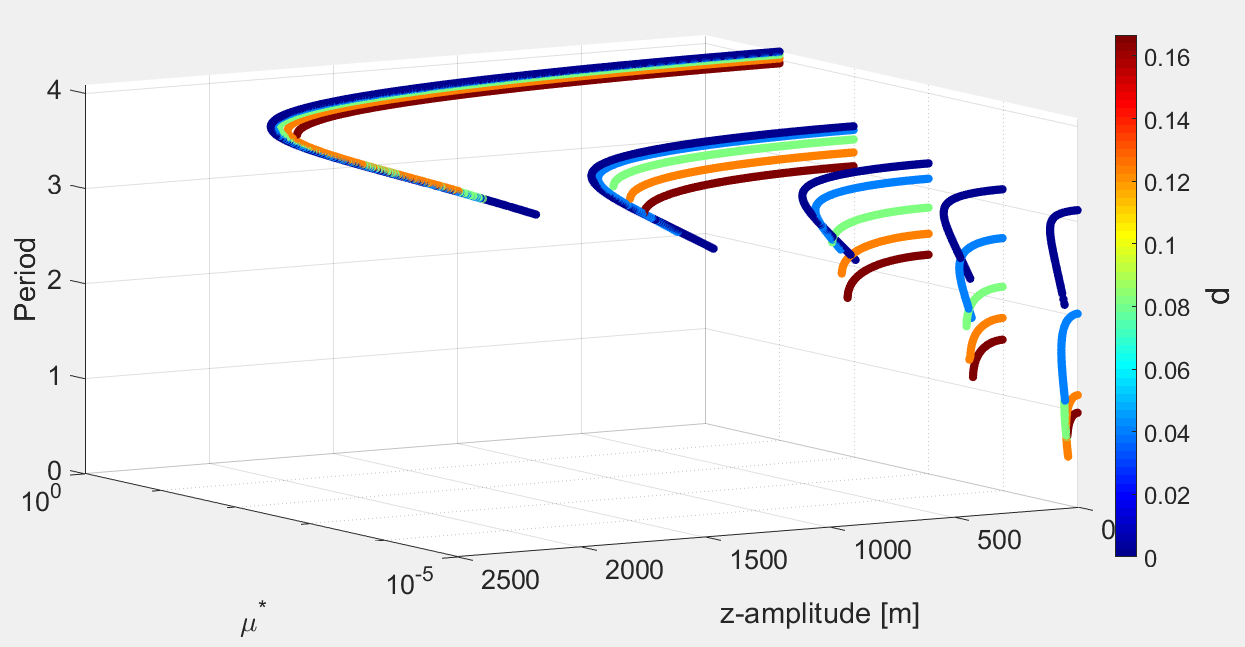}
	\caption{Period of the halo orbits around $ L_2 $ as a function of $ d $ and $ \mu^* $.}
	\label{Periodhalol2}
\end{figure}

Now we analyze how the period of the halo orbits around $ L_1 $ and $ L_2 $ is affected by $ d $ and $ \mu^* $. As $ d $ increases and $ \mu^* $ is kept constant, the periods of the halo orbits decrease, as shown in Figures \ref{Periodhalo} and \ref{Periodhalol2}.

This is because the gravitational attraction is stronger near the particle, due to the mass distribution of the secondary body, causing the acceleration to increase and the orbital period to decrease. As the amplitude of the halo orbit increases, its orbital period becomes shorter.

Considering the elongated shape of the asteroid, but keeping $ d $ constant and increasing $ \mu^* $, we notice that the period of the halo orbits become longer. This is because, as $ \mu^* $ increases, the equilibrium point move away from the secondary body, thus the halo orbits are further away from the secondary body, which causes the gravitational acceleration to decrease, and thus the orbital period of the particle along the orbit to increase.

\begin{figure}[!htbp]
	\centering\includegraphics[scale=0.33]{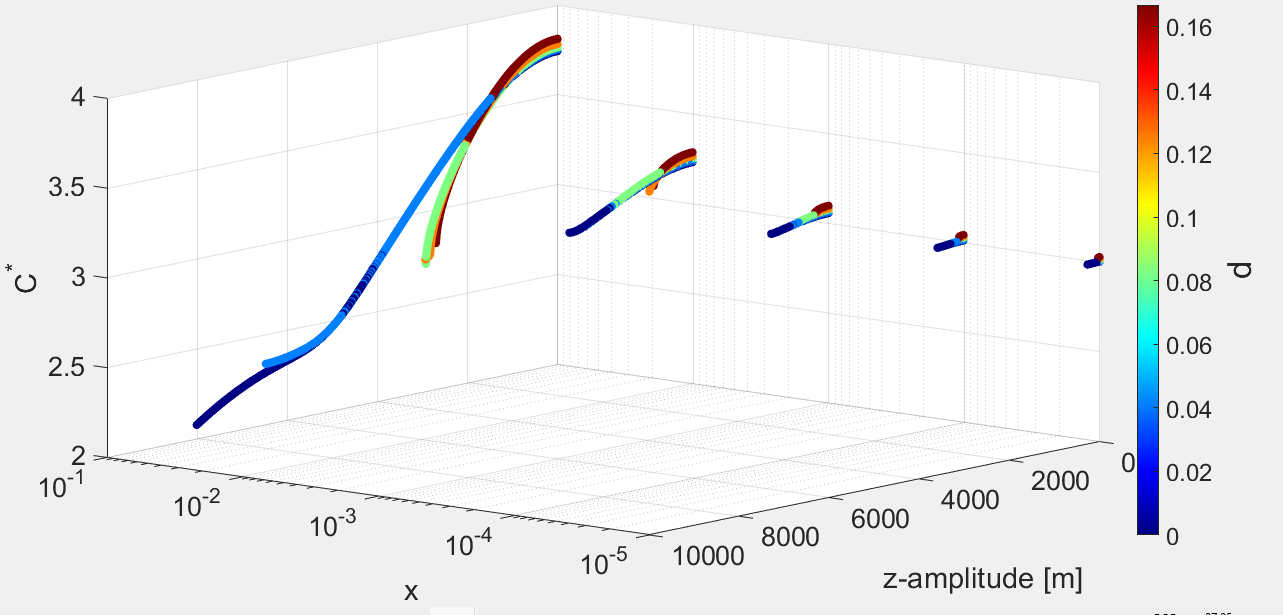}
	\caption{Jacobi constant of the halo orbits around $ L_1 $ with respect to $ d $ and $ \mu^* $.}
	\label{C_halo_l1}
\end{figure}

\begin{figure}[!htbp]
	\centering\includegraphics[scale=0.33]{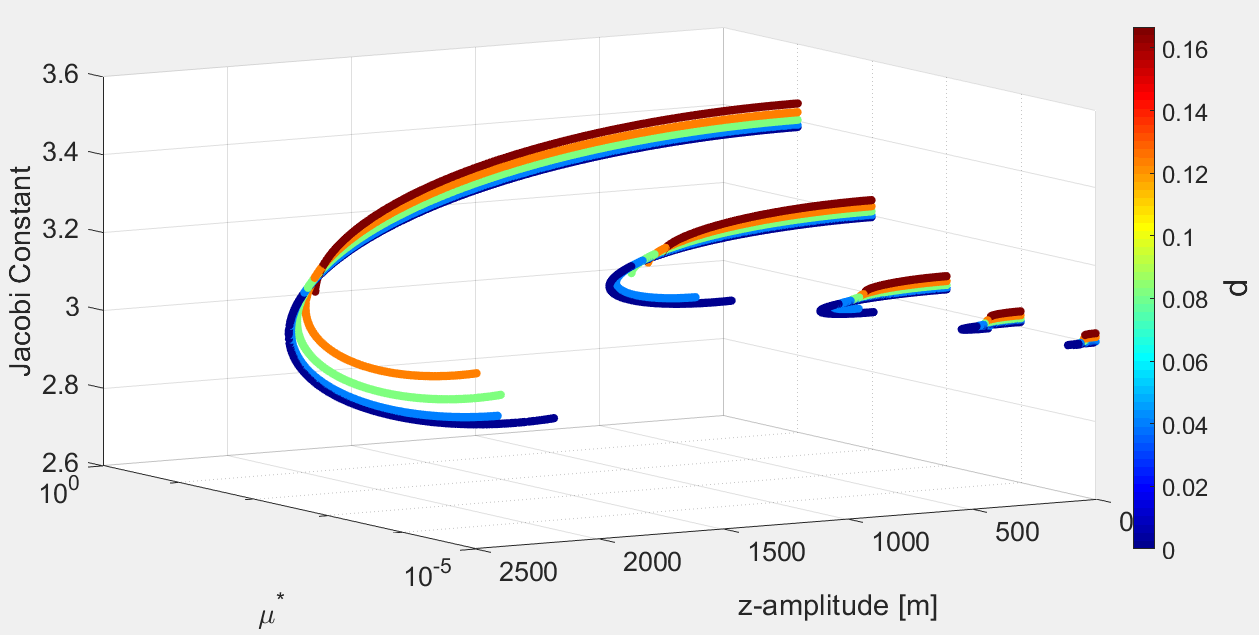}
	\caption{Jacobi constant of the halo orbits around $ L_2 $ with respect to $ d $ and $ \mu^* $.}
	\label{C_halo_l2}
\end{figure}

Figures \ref{C_halo_l1} and \ref{C_halo_l2} provide information on the behavior of the Jacobi constant of the halo orbits as a function of $ d $ and $ \mu^* $. Note that when $ d $ or $ \mu^* $ increases, the range of value of the Jacobi constant also increases. This is important information in terms of the application of space mission. Note that the larger the mass ratio of the system, or the longer the secondary body, less energy is needed for the halo orbits to branch from the planar orbits.

\section{Conclusion}
\label{conclusion}

In this paper, the general dynamical environment in the vicinity of binary asteroid systems is explored. Based on the physical and orbital parameters of type A asteroids, the positions of the collinear balance points as a function of angular velocity were computed. We found that the locations of the collinear equilibrium points $ L_3 $ and $ L_2 $ are more sensitive to changes in the rotation rate, compared to $ L_1 $.

Families of planar and Halo orbits were computed around these equilibrium points and we found that the closer the periodic orbits are to the equilibrium point, the more unstable they are.

Numerical evidence shows that the stability of the periodic orbits around the equilibrium points depends on the size of the secondary body and the mass ratio of the system. We observed that, the more elongated the secondary body, the more unstable the planar orbits are. Additionally, we detected unstable and stable halo orbits when $ d $ = 0 and when $ d ~ \neq ~ 0 $.

Finally, we observed that, keeping the mass ratio constant, the more elongated the secondary body, the lower the orbital periods of planar and halo orbits around the equilibrium points.

Thus, if a spacecraft were to be placed in the vicinity of an equilibrium point, fuel consumption required for orbital maintenance would be higher around more elongated secondary bodies.

\section{Acknowledgements}

The authors wish to express their appreciation for the support provided by: grants 140501/2017-7, 150678/2019-3, 422282/2018-9 and 301338/2016-7 from the National Council for Scientific and Technological Development (CNPq); grants 2016/24561-0, 2016/18418-0, 2018/06966-8 and 2018/07377-6 from S\~ao Paulo Research Foundation (FAPESP); grant 88887.374148/2019-00 from the National Council for the Improvement of Higher Education (CAPES); grant E-26/201.877/2020, from Rio de Janeiro Research Foundation (FAPERJ) and to the National Institute for Space Research (INPE). This publication has been supported by the RUDN University Scientific Projects Grant System, project No 202235-2-000


\begin{thebibliography}

\bibitem[Aljbaae et al.(2017)]{aljbaae_2017} Aljbaae, S., Chanut, T.~G.~G., Carruba, V., et al.\ 2017, \mnras, 464, 3552. doi:10.1093/mnras/stw2619

\bibitem[Aljbaae et al.(2020)]{2020MNRAS.496.1645A} Aljbaae, S., Prado, A.~F.~B.~A., Sanchez, D.~M., et al.\ 2020, \mnras, 496, 1645. doi:10.1093/mnras/staa1634.

\bibitem[Barbosa Torres dos Santos et al.(2017)]{2017ApSS.362...61B} Barbosa Torres dos Santos, L., Bertachini de Almeida Prado, A.~F., \& Merguizo Sanchez, D.\ 2017, \apss, 362, 61. doi:10.1007/s10509-017-3030-2.

\bibitem[Blesa.(2006)]{Blesa} Blesa, F.\ 2006, Monografías del Seminario Matemático García de Galdeano, 33, 67.

\bibitem[Bosanac(2016)]{natasha2} Bosanac, N.\ 2016, Ph.D. Thesis.

\bibitem[Broucke(1969)]{1969AIAAJ...7.1003B} Broucke, R.\ 1969, AIAA Journal, 7, 1003. doi:10.2514/3.5267

\bibitem[Celik \& Sanchez (2017)]{Celik2017} Celik, O., Sanchez, J.P., \ 2017, Journal of Guidance, Control, and Dynamics 40, 1390-1420. doi:10.2514/1.G002181.

\bibitem[Chanut et al.(2015)]{chanut_2015a} Chanut, T.~G.~G., Aljbaae, S., \& Carruba, V.\ 2015, \mnras, 450, 3742. doi:10.1093/mnras/stv845.

\bibitem[Chappaz \& Howell(2015)]{Chappaz} Chappaz, L. \& Howell, K.~C.\ 2015, Celestial Mechanics and Dynamical Astronomy, 123, 123. doi:10.1007/s10569-015-9632-5.

\bibitem[de Almeida Junior \& Prado (2022)]{dealmeidajrbert22} de Almeida Junior, A.K., Prado, A.F.B.A.\ 2022, Scientific Reports 12, 4148 (2022). doi:10.1038/s41598-022-08046-x

\bibitem[dos Santos et al.(2017)]{2017Ap&SS.362..202D} dos Santos, L.~B.~T., de Almeida Prado, A.~F.~B., \& Sanchez, D.~M.\ 2017, \apss, 362, 202. doi:10.1007/s10509-017-3177-x.

\bibitem[dos Santos et al.(2020)]{leotripole} dos Santos, L.~B.~T., Marchi, L., Sousa-Silva, P.~A., et al.\ 2020, rmxaa, 56, 269. doi:10.22201/ia.01851101p.2020.56.02.09

\bibitem[Feng et al.(2016)]{2016AdSpR..58..387F} Feng, J., Noomen, R., Visser, P., et al.\ 2016, Advances in Space Research, 58, 387. doi:10.1016/j.asr.2016.04.032.

\bibitem[Ferrari et al. (2016)]{ferrarilav2016} Ferrari, F., Lavagna, M., Howell, K. \ 2016, Celestial Mechanics and Dynamical Astronomy 125 (4), 413-433. doi:10.1007/s10569-016-9688-x.

\bibitem[Grebow(2006)]{Grebow} Daniel J. Grebow \ 2009, Master of Science in Aeronautics and Astronautics Thesis, Purdue University.

\bibitem[Howell(1982)]{Howell1} Howell, K.~C.\ 1982, Astrodynamics 1981, 528.

\bibitem[Haapala et al.(2015)]{Haapala} Haapala, A.~F., Howell, K.~C., \& Folta, D.~C.\ 2015, Acta Astronautica, 112, 1. doi:10.1016/j.actaastro.2015.02.024.

\bibitem[Jacobson \& Scheeres(2011)]{2011IcarJacobson} Jacobson, S.~A. \& Scheeres, D.~J.\ 2011, icarus, 214, 161. doi:10.1016/j.icarus.2011.04.009

\bibitem[Lan et al.(2017)]{2017Ap&SS.362..169L} Lan, L., Yang, H., Baoyin, H., et al.\ 2017, \apss, 362, 169. doi:10.1007/s10509-017-3148-2.

\bibitem[Liu et al.(2011)]{2011Ap&SS.333..409L} Liu, X., Baoyin, H., \& Ma, X.\ 2011, \apss, 333, 409. doi:10.1007/s10509-011-0669-y.

\bibitem[Margot et al.(2015)]{2015aste.book..355M} Margot, J.-L., Pravec, P., Taylor, P., et al.\ 2015, Asteroids IV, 355. doi:10.2458/azu\_uapress\_9780816532131-ch019.

\bibitem[McCuskey(1963)]{mc} McCuskey, S.~W.\ 1963, Reading, Mass., Addison-Wesley Pub. Co. [1963].

\bibitem[Meyer \& Hall(1992)]{Kenneth} Meyer, K.~R. \& Hall, G.~R.\ 1992, Science, 255, 1756.

\bibitem[Pravec et al.(2006)]{2006Icar..181...63P} Pravec, P., Scheirich, P., Ku{\v{s}}nir{\'a}k, P., et al.\ 2006, icarus, 181, 63. doi:10.1016/j.icarus.2005.10.014.

\bibitem[Pravec \& Harris(2007)]{2007Icar..190..250P} Pravec, P. \& Harris, A.~W.\ 2007, icarus, 190, 250. doi:10.1016/j.icarus.2007.02.023.

\bibitem[Pravec et al.(2016)]{2016Icar267PRAVEC} Pravec, P., Scheirich, P., Ku{\v{s}}nir{\'a}k, P., et al.\ 2016, icarus, 267, 267. doi:10.1016/j.icarus.2015.12.019

\bibitem[Riaguas et al.(1999)]{1999imda.coll..169R} Riaguas, A., Elipe, A., \& Lara, M.\ 1999, Impact of Modern Dynamics in Astronomy, 169.

\bibitem[Riaguas et al.(2001)]{2001CeMDA..81..235R} Riaguas, A., Elipe, A., \& L{\'o}pez-Moratalla, T.\ 2001, Celestial Mechanics and Dynamical Astronomy, 81, 235.

\bibitem[Santos et al.(2021)]{leotripole3d} Santos, L.~B.~T., Marchi, L.~O., Aljbaae, S., et al.\ 2021, \mnras, 502, 4277. doi:10.1093/mnras/stab198

\bibitem[Scheeres et al.(2021)]{ScheeresBAresi2019} Scheeres, D.J., Van wal, S., Olikara, Z., Baresi, N.\ 2019, Advances in Space Research 63, 476-495.
doi:10.1016/j.asr.2018.10.016

\bibitem[Szebehely(1967)]{1967torp.book.....S} Szebehely, V.\ 1967, New York: Academic Press, |c1967.

\bibitem[Tardivel \& Scheeres (2013)]{TardivelScheeres2013} Tardivel, S., Scheeres, D.J., 2013, Journal of Guidance, Control and Dynamics 36, 700-709. doi:10.2514/1.59106.

\bibitem[Walsh et al.(2008)]{2008Natur.454..188W} Walsh, K.~J., Richardson, D.~C., \& Michel, P.\ 2008, \nat, 454, 188. doi:10.1038/nature07078.

\bibitem[Walsh \& Jacobson(2015)]{2015aste.book..375W} Walsh, K.~J. \& Jacobson, S.~A.\ 2015, Asteroids IV, 375. doi:10.2458/azu\_uapress\_9780816532131-ch020.

\bibitem[Wang et al.(2017)]{2017Ap&SS.362..229W} Wang, W., Yang, H., Zhang, W., et al.\ 2017, \apss, 362, 229. doi:10.1007/s10509-017-3206-9.

\bibitem[Wen \& Zeng (2022)]{WenZeng} Wen, T., Zeng, X.\ 2022, Advances in Space Research, 69, 2223. doi:10.1016/j.asr.2021.12.021.

\bibitem[Werner(1994)]{1994CeMDA..59..253W} Werner, R.~A.\ 1994, Celestial Mechanics and Dynamical Astronomy, 59, 253. doi:10.1007/BF00692875.

\bibitem[Yang et al.(2015)]{2015RAA....15.1571Y} Yang, H.-W., Zeng, X.-Y., \& Baoyin, H.\ 2015, Research in Astronomy and Astrophysics, 15, 1571. doi:10.1088/1674-4527/15/9/013.

\bibitem[Yang et al.(2017)]{2017Ap&SS.362...27Y} Yang, H., Baoyin, H., Bai, X., et al.\ 2017, \apss, 362, 27. doi:10.1007/s10509-017-3007-1.

\bibitem[Zeng et al.(2015)]{2015Ap&SS.356...29Z} Zeng, X., Jiang, F., Li, J., et al.\ 2015, \apss, 356, 29. doi:10.1007/s10509-014-2187-1.

\bibitem[Zeng et al.(2016)]{2016JGCD...39.1223Z} Zeng, X., Gong, S., Li, J., et al.\ 2016, Journal of Guidance Control Dynamics, 39, 1223. doi:10.2514/1.G001061.

\bibitem[Zeng et al.(2016)]{2016Ap&SS.361...14Z} Zeng, X., Baoyin, H., \& Li, J.\ 2016, \apss, 361, 14. doi:10.1007/s10509-015-2598-7

\bibitem[Zeng et al.(2018)]{2018AJ....155...85Z} Zeng, X., Zhang, Y., Yu, Y., et al.\ 2018, \aj, 155, 85. doi:10.3847/1538-3881/aaa483

\bibitem[Zhang et al.(2020)]{2020AcAau.177...15Z} Zhang, R., Wang, Y., Shi, Y., et al.\ 2020, Acta Astronautica, 177, 15. doi:10.1016/j.actaastro.2020.07.006.

\bibitem[Zimovan-Spreen et al.(2020)]{2020CeMDA.132...28Z} Zimovan-Spreen, E.~M., Howell, K.~C., \& Davis, D.~C.\ 2020, Celestial Mechanics and Dynamical Astronomy, 132, 28. doi:10.1007/s10569-020-09968-2
\end{thebibliography}
\end{document}